\documentclass[preprint,pre,oneside,superscriptaddress,floatfix]{revtex4-1} 

\usepackage{graphicx}
\usepackage{color}
\usepackage{placeins}

\usepackage{amsmath}
\usepackage{amssymb}
\usepackage{mathtools}
\usepackage{bbold}

\usepackage{siunitx}
\DeclareSIUnit\Molar{\textsc{M}}

\usepackage[]{hyperref}

\newcolumntype{K}{>{\centering\arraybackslash$}p{1cm}<{$}}

\renewcommand{\vec}[1]{\boldsymbol{#1}}

\def\epsilon{\varepsilon}
\def\theta{\vartheta}
\def\rho{\varrho}
\def\Int#1#2#3{\int_{#1}\!\mathrm{d}^{#2}{#3}\;}

\begin{document}

\title{Heterogeneous surface charge confining an electrolyte solution}

\date{\today}

\author{Maximilian Mu{\ss}otter}
\email{mussotter@is.mpg.de}
\affiliation{
	Max Planck Institute for Intelligent Systems, 
	Heisenbergstr.\ 3,
	70569 Stuttgart,
	Germany
}
\affiliation{
	IV$^{\text{th}}$ Institute for Theoretical Physics, 
	University of Stuttgart,
	Pfaffenwaldring 57,
	70569 Stuttgart,
	Germany
}
\author{Markus Bier}
\email{bier@is.mpg.de}
\affiliation{
	Max Planck Institute for Intelligent Systems, 
	Heisenbergstr.\ 3,
	70569 Stuttgart,
	Germany
}
\affiliation{
	IV$^{\text{th}}$ Institute for Theoretical Physics, 
	University of Stuttgart,
	Pfaffenwaldring 57,
	70569 Stuttgart,
	Germany
}
\affiliation{
	Fakult\"at Angewandte Natur- und Geisteswissenschaften,
	University of Applied Sciences W\"urzburg-Schweinfurt,
	Ignaz-Sch\"on-Str. 11,
	97421 Schweinfurt,
	Germany
}
\author{S. Dietrich}
\affiliation{
	Max Planck Institute for Intelligent Systems, 
	Heisenbergstr.\ 3,
	70569 Stuttgart,
	Germany
}
\affiliation{
	IV$^{\text{th}}$ Institute for Theoretical Physics, 
	University of Stuttgart,
	Pfaffenwaldring 57,
	70569 Stuttgart,
	Germany
}

\begin{abstract}
	The structure of dilute electrolyte solutions close to a surface carrying a spatially inhomogeneous surface charge distribution is investigated by means of classical density functional
	theory (DFT) within the approach of fundamental measure theory (FMT). For electrolyte solutions the influence of these inhomogeneities is particularly strong because the corresponding characteristic length scale is the 
	Debye length, which is large compared to molecular sizes. Here a fully three-dimensional investigation is performed, which accounts explicitly for the solvent particles, and thus provides insight into effects caused by 
	ion-solvent coupling. The present study introduces a versatile framework to analyze a broad range of types of surface charge heterogeneities even beyond the linear response regime. This reveals a sensitive dependence
	of the number density profiles of the fluid components and of the electrostatic potential on the magnitude of the charge as well as on details of the surface charge patterns at small scales.
\end{abstract}

\maketitle

\section{Introduction}

In a wide spectrum of research areas and applications --- ranging from electrochemistry
\cite{Bagotsky2006, Schmickler2010} and wetting phenomena \cite{Dietrich1988, Schick1990} via
coating \cite{Wen2017} and surface patterning \cite{Vogel2012, Nee2015} to colloid science
\cite{Russel1989, Hunter2001, Bakhshandeh2019} and microfluidics \cite{Lin2011, GalindoRosales2018} --- there is a significant interest
in understanding the structure of electrolyte solutions at solid substrates. Most of the theoretical
studies dealing with such fluids close to a substrate neglect heterogeneities in the interaction between the 
wall and the fluid, modeling the substrate as being uniform. On one hand this approach simplifies the calculations significantly whereas on the other hand
there is a lack of experimental data concerning the actual local structure of these fluids near substrates. In the case of \emph{electrically neutral} fluids and \emph{uncharged} walls
this simplification is typically well justified because, besides wetting transitions, the bulk correlation length sets
the length scale, on which heterogeneities of surface properties influence the fluid \cite{Andelman1991}. This bulk correlation length
is, sufficiently far from critical points, of the order of a few molecular diameters, rendering any heterogeneity to be of negligible importance. In contrast to this short length scale,
a dilute electrolyte solution close to nonuniformities of the \emph{surface charge density} of a charged substrate is influenced on the length scale
of the Debye length, which is, for this type of solutions, much larger than the size of the fluid constituents. Additionally, surface charge
heterogeneities of typical substrates (e.g., minerals and polyelectrolytes) are usually also of the order of the Debye length of the fluid close to these substrates \cite{Chen2005, Chen2006, Chen2009}.
Consequently, for the treatment of dilute electrolyte solutions in contact with charged surfaces, the approximation of assuming uniform surface charge densities is questionable.

Over the last years, an increasing interest in these surfaces has developed, leading to a number of studies investigating the influence of heterogeneously charged walls, for example on the
effective interaction between two substrates in contact with an electrolyte solution \cite{Naji2010, Ben-Yaakov2013, Naji2014, Bakhshandeh2015, Ghodrat2015a, Ghodrat2015b, Adar2016,
Adar2017a, Adar2017b, Ghosal2017, Zhou2017}. These studies revealed the effective interaction in case of nonuniform substrates to cause lateral forces in addition to the ones in normal direction, which are commonly known.
Despite describing the solute components in a wide range of ways, all those studies neglect the size of the solvent particles and its influence on the permittivity of the fluid,
treating it as structureless dielectric continuum. As has been shown in previous studies \cite{Onuki2004,Bier2012a, Bier2012b}, due to a competition between the solvation and the
electrostatic interaction, there are coupling effects occurring in bulk electrolyte solutions, which cannot be captured by these simple approaches. However, in particular in the presence of ion-solvent
coupling, fluctuations of the solvent density decay on the scale of the Debye length, which leads to inhomogeneities in the wall-solvent interaction influencing the structure of the electrolyte
solution in contact with the charged substrate on a length scale much larger than molecular sizes. A very recent example of such a study, deriving exact solutions of the shape of the electrostatic potential in an
electrolyte solution close to a heterogeneously charged surface within Poisson-Boltzmann theory, is given by Ref.~\cite{Samaj2019}. There, previous work of the present authors \cite{Mussotter2018} was expanded with
respect to the description of non-linear responses, albeit not explicitly including the solvent and neglecting the spatial extent of all fluid particles.
Moreover, in Ref. \cite{Gillespie2017} a one-dimensional wall with a single, isolated step in the surface charge in contact with a hard-sphere electrolyte solution was studied in a broad parameter range, concerning both surface specifications and characteristics of the electrolyte solution. It was shown, that the valences of the ions, their respective sizes, their concentration, and the strength of the surface charge can lead to various structural effects in the fluid structure both perpendicular and parallel to the wall. However, in that study, the solvent was again treated only implicitly and thus coupling effects have been neglected.

In the present analysis, we aim for a deeper understanding of the structural effects of surface charge nonuniformities on a nearby dilute electrolyte solution in terms of all fluid components.
The system is studied by means of density functional theory (DFT) in combination with fundamental measure theory (FMT), which has been shown to be a powerful framework for investigating fluid
structures in terms of density profiles \cite{Evans1979, Evans1990, Evans1992}. The study at hand is concerned with explicitly calculating the structure of an electrolyte solution composed of neutral solvent particles
and a single univalent salt component, described as hard spheres. As for the structure of the two-dimensional surface nonuniformities, they can be arbitrary in strength and also their 
spatial arrangement can de facto be chosen freely, with the computational capacities being the only limitation. However, here we restrict ourselves to periodic surface charge patterns.
Furthermore, we lift the constraint of overall charge neutral walls, as has been used in Refs.~\cite{Naji2010, Ghodrat2015a, Adar2016, Zhou2017}.
The present study addresses the open questions from Ref. \cite{Mussotter2018} concerning the influence of microscopic details and non-linear effects on the structure of a dilute electrolyte solution close to heterogeneously
charged walls.
In the present study we have chosen a small subset of the parameter range analyzed in  Ref. \cite{Gillespie2017}, for which it has been shown, that the valences have a negligible effect and that the width of the region, which is influenced by a variation of the surface charge, is computationally manageable (see Sec. \ref{Sec:Parameters}). Furthermore, we have focussed on the effect of multiple heterogeneities of the sort of the ones discussed in Ref. \cite{Gillespie2017}, thereby creating a two-dimensional patterned surface.

In the following, first the setup and the formalism will be introduced in Sec.~\ref{Sec:Formalism}. Secondly, in Sec.~\ref{Sec:ResultsDiscussion} various selected surface charge patterns 
are studied. From a simple homogeneously charged surface investigated in Sec.~\ref{Sec:Constant}, we move towards more complicated charge distributions such as a sinusoidal shape (Sec.~\ref{Sec:Sinus}), patch-like, or 
rectangular patterns (Sec.~\ref{Sec:VariousPatterns}).
Conclusions and a summary are presented in Sec.~\ref{Sec:ConclusionsSummary}.

\section{Theoretical foundations\label{Sec:Formalism}}

\subsection{Setup\label{Sec:Setup}}

In the present paper, the interplay of an electrically nonuniform hard wall and a fluid, comprising hard spheres and monovalent ions, is studied.
For $z<0$ the system consists of a semi-infinite impenetrable (hard) wall; on the other hand for $\mathcal{V} = \{\vec{r}\in\mathbb{R}^3|z>0\}$ the system is occupied by a fluid mixture of hard spheres,
where $z$ is one of the three spatial coordinates $\vec{r}=(x,y,z)$.
This fluid is an electrolyte solution with three particle species: an uncharged solvent (index "1"), monovalent cations (index "2"), and monovalent anions (index "3"),
all of which are taken to be of the same size.
All these interactions are either of electrostatic nature, as caused by electric monopoles at the surface of the wall ($z=0$) and by the monovalent ions, or of a nonelectrostatic,
purely repulsive nature caused by the steric repulsion of the hard spheres and the hard wall.
The precise geometries of the electric nonuniformities at the wall are given in the context of the various scenarios studied in the subsequent Sec.~\ref{Sec:ResultsDiscussion}.

However, in all cases we assume the nonuniformities at the wall to be periodic with a unit cell of size $L_x\times L_y\eqqcolon P_xR_1\times P_yR_1$, where $R_1$ is the radius
of the solvent particles. $P_x$ and $P_y$ are the dimensionless widths of the box, for which the numerical evaluations are performed (see Appendix \ref{Sec:AppDisc}). Furthermore, we assume that all deviations from the bulk behavior are located in close proximity to the wall, justifying a restriction of the numerical treatment
to a length $L_z = P_zR_1$ in the direction normal to the wall and assuming the densities to take on their respective bulk values for $z>L_z$. With the same justification we assume
the electrostatic potential $\Psi$ to decay purely exponentially with the decay length given by the Debye length $1/\kappa = \sqrt{\frac{\epsilon_r}{8\pi l_B I}}$ for $z>L_z$, where
$l_B=\frac{e^2}{4\pi\epsilon_0 k_{\mathrm{B}}T}=\SI{56.8}{nm}$ is the vacuum Bjerrum length, $\epsilon_r$ is the relative permittivity, and $I = \rho_{2,\text{b}} = \rho_{3,\text{b}}$ is the ionic strength, which, in the present
case of monovalent ions, equals the bulk number density of the cations and the anions.

In order to tackle the system described above numerically, we introduce a discretization of the system by using discrete versions of the equations derived in the following. However, in order to
illustrate the approach, we use the continuous expressions. For further details concerning the discretization see Appendix \ref{Sec:AppDisc}.

\subsection{Density functional theory \label{Sec:DFT}}

In order to determine the equilibrium number density profiles $\vec{\rho}=(\rho_1,\rho_2,\rho_3)$ of the three species we use density functional theory \cite{Evans1979, Evans1990, Evans1992}.
To this end we establish the approximate density functional $\beta\Omega[\vec{\rho}=(\rho_1,\rho_2,\rho_3)]$, which is minimized by the equilibrium number density profiles of the three species.

The exactly known expression for the ideal gas contribution
\begin{equation}
	\beta\Omega_{\mathrm{id}}[\vec{\rho}] = \Int{\mathcal{V}}{3}{r} \sum_{i}\left(\rho_i (\ln(\rho_i\Lambda_i^3) - 1 - \beta\mu_i)\right),
\end{equation}
where $\beta = 1/(k_{\mathrm{B}}T)$ is the inverse thermal energy, $\mu_i$ are the chemical potentials of the three species, and $\Lambda_i$
are the thermal de Broglie wavelengths of the three species. In addition to this, we use fundamental measure theory (FMT) in the White Bear I version \cite{Roth2002} to account for
the hard sphere nature of the fluid constituents. 
This leads to an excess free-energy functional for the hard core part given by
\begin{equation}
	\beta F^{\mathrm{hc}}[\vec{\rho}] = \Int{\mathcal{V}}{3}{r'} \Phi(\{n_{\alpha}(\vec{r}')\})\label{eq:fmt},
\end{equation}
with the volume $\mathcal{V}=\mathbb{R}^2\times(0,\infty)$ and the reduced free-energy density 
\begin{equation}
	\Phi = -n_0\ln(1-n_3) + \frac{n_1n_2 - \vec{n}_1\cdot\vec{n}_2}{1-n_3} + (n^3_2 - 3n_2\vec{n}_2\cdot\vec{n}_2)\frac{n_3 + (1-n_3)^2\ln(1-n_3)}{36\pi n^2_3(1-n_3)^2}
\end{equation}
as a function of the weighted densities
\begin{equation}
	n_{\alpha}(\vec{r}) = \sum_{i=1}^{3}\Int{\mathcal{V}}{3}{r'}\rho_i(\vec{r}-\vec{r}')\omega^{(\alpha)}_i(\vec{r}').
\end{equation}
Following Ref. \cite{Roth2002}, the weight functions are given by
\begin{align}
	\omega^{(3)}_i(r) &= \Theta(R_i - r),\\
	\omega^{(2)}_i(r) &= \delta(R_i - r),\\
	\omega^{(1)}_i(r) &= \frac{\omega^{(2)}_i(r)}{4\pi R_i},\\
	\omega^{(0)}_i(r) &= \frac{\omega^{(2)}_i(r)}{4\pi R^2_i},\\
	\vec{\omega}^{(2)}_i(r) &= \frac{\vec{r}}{r}\delta(R_i - r), \text{ and}\\
	\vec{\omega}^{(1)}_i(r) &= \frac{\vec{\omega}^{(2)}_i(r)}{4\pi R_i}\label{eq:fmt2},
\end{align}
where $R_i$ are the radii of the three species denoted by index $i$.

Furthermore, in the present study electrostatic contributions play a role, all of which are combined in the form of the electric field energy density $\beta U_{\text{el}}[\vec{\rho}]$ given by (see
Eqs. \eqref{eq:EUndUelConnection} and \eqref{eq:appendixUE})
\begin{equation}
	\beta  U_{\text{el}}[\vec{\rho}] = \frac{\beta}{2}\Int{\mathcal{V}}{3}{r} \epsilon_0\epsilon_r(\vec{r})(\nabla\Psi(\vec{r}))^2.
	\label{eq:betaU}
\end{equation}
Here, $\epsilon_0$ is the vacuum permittivity, $\epsilon_r$ is the relative permittivity, and $\Psi$ is the electrostatic potential profile. The
relative permittivity $\epsilon_r$ will be  discussed in more detail in Sec. \ref{Sec:Epsilon}. Furthermore, it should be noted, that the calculation of the electrostatic field energy is mathematically identical to other commonly used variants, such as the ones in Refs. \cite{Gillespie2003, Gillespie2005} (see Appendix \ref{Sec:ApUel}). In the context of the present investigation the expression in Eq. \eqref{eq:betaU} is chosen due to its close connection to the applied numerical two-step minimization method. Its reliability can readily be tested by comparing the results for a homogeneous wall (Sec. \ref{Sec:Constant}) with corresponding results in Refs. \cite{Outhwaite2004, Outhwaite2007, Gillespie2018}.
The resulting expression for the density functional used in the present study is
\begin{equation}
	\beta\Omega[\vec{\rho}] = \Int{\mathcal{V}}{3}{r} \left(\sum_{i}\rho_i (\ln(\rho_i\Lambda_i^3) - 1 - \beta\mu_i)\right) + \beta F^{\text{hc}}[\vec{\rho}] + \beta U_{\text{el}}[\vec{\rho}]\label{eq:DFT},
\end{equation}
which in turn can now be used to determine the Euler-Lagrange equations. Whereas the hard-core contribution $\beta F^{\text{hc}}[\vec{\rho}]$ to the density functional in Eq. \eqref{eq:DFT} is based on the fundamental measure theory (FMT) described by Eqs. \eqref{eq:fmt} - \eqref{eq:fmt2}, the electrostatic contribution $\beta U_{\text{el}}[\vec{\rho}]$ in Eq. \eqref{eq:betaU} is a random-phase approximation (RPA). Hence, one cannot expect that the second-moment Stillinger-Lovett sum rule or the consistency between the test-particle and the Ornstein-Zernike route to the pair distribution functions are fulfilled. However, this should not be regarded as a serious disadvantage, because the present investigation is addressing the fluid structure close to a solid surface, which is strongly dominated by the influence of the external field of the wall and is influenced only to a small extent by correlations from the bulk.

\subsection{Derivation of the Euler-Lagrange equations\label{Sec:ELG}}
Using the previously established density functional (see Sec.~\ref{Sec:DFT}) we now can derive the corresponding Euler-Lagrange equations, which provide the equilibrium density profiles.

Since minimizing with respect to all three density profiles and the potential distribution at the same time is computationally very costly, we divide up the minimization in first minimizing
for the equilibrium form $\Psi(\vec{r})$ of $\psi(\vec{r})$ at fixed density profiles $\vec{\rho}$ and subsequently minimizing with respect to the three density profiles $\vec{\rho}(\vec{r})$.

In order to determine the equilibrium form of $\psi(\vec{r})$ for Eq. \eqref{eq:betaU}, we introduce 
\begin{equation}
	\mathcal{E}[\psi,q,\epsilon_r,\sigma] = \Int{\mathcal{V}}{3}{r}\left(\frac{\epsilon_0\epsilon_r(\vec{r})}{2}(\nabla\psi(\vec{r}))^2-q(\vec{r})\psi(\vec{r})\right)-\Int{\partial\mathcal{V}}{2}{s}\sigma(\vec{s})\psi(\vec{s},0),
	\label{eq:Edef}
\end{equation}
where $q(\vec{r})=e(\rho_2(\vec{r})-\rho_3(\vec{r}))$ is the local charge density, and $\sigma(\vec{s})$ is the surface charge density at the wall surface $\partial\mathcal{V}=\{\vec{r}\in\mathbb{R}^3|(x,y,z=0) = (\vec{s},0)\}$.
As shown in Appendix \ref{Sec:ApUel}, the electrostatic field energy can be written as
\begin{equation}
	\beta  U_{\text{el}}[\vec{\rho}] = -\beta\mathcal{E}[\Psi,q,\epsilon_r,\sigma],
\end{equation}
that is, given the equilibrium potential $\Psi(\vec{r})$ for a given charge distribution $q(\vec{r})$, the electrostatic contribution to the density functional
can be expressed in terms of $\mathcal{E}[\Psi,q,\epsilon_r,\sigma]$.

Additionally, by construction the variation of $\mathcal{E}[\psi,q,\epsilon_r,\sigma]$ with respect to $\psi$ vanishes for the equilibrium profile $\Psi$ (see Appendix \ref{Sec:ApE}).
Therefore, we can find the equilibrium potential distribution $\Psi$ corresponding to a given distribution of the densities, i.e., $\sigma(\vec{s})$, $q(\vec{r})$, and $\epsilon_r(\vec{r})$, by minimizing $\mathcal{E}$
with respect to $\psi$. Following these lines and incorporating the numerically necessary discretization as outlined in Sec.~\ref{Sec:Setup}, one finds the Euler-Lagrange equations (Eqs. \eqref{eq:Psi0} - \eqref{eq:PsiN}) for
the electrostatic potential, which depend on the distance from the wall.

After determining the electrostatic potential distribution by solving Eqs. \eqref{eq:Psi0}, \eqref{eq:Psik}, and \eqref{eq:PsiN}, we can use the resulting solution for $\psi(\vec{r})=\Psi(\vec{r})$ in accordance with Eq. \eqref{eq:betaU}
in order to determine the density functional and the following Euler-Lagrange equations for the three number densities $\vec{\rho}$:
\begin{align}
	\ln(|\mathcal{C}^*|\rho_j(\vec{r})) = \mu_j^*-\sum_{\alpha}\Int{\mathcal{V}}{3}{r'} p_{\alpha}\frac{\partial\Phi(\vec{r})}{\partial n^{\alpha}(\vec{r'})} \omega^{(\alpha)}_j(\vec{r'} - \vec{r}) + \frac{\partial\beta\mathcal{E}}{\partial\rho_j(\vec{r})}. \label{eq:ELGDENS}
\end{align}
Here $|\mathcal{C}^*|$ is the size of one of the cells used in the numerical implementation (see Appendix \ref{Sec:AppDisc}), $\mu_j^*=\beta\mu_j - \ln(\Lambda_j^3/|\mathcal{C}^*|)$ is the dimensionless effective chemical potential of species $j$, and the prefactor is $p_{\alpha}=-1$ for vectorial
weights $\vec{\omega}^{(\alpha)}$ and $p_{\alpha}=1$ for scalar weights. The three terms result from the ideal gas contribution, the FMT contribution, and the electrostatic interactions, respectively.

\subsection{Choice of parameters \label{Sec:Parameters}}
For solving the Euler-Lagrange equations obtained above (Sec.~\ref{Sec:ELG}) and in Appendix \ref{Sec:ApE}, there are certain parameters which have to be fixed in advance. Besides the surface charge distribution
$\sigma(\vec{s})$, which is varied for each calculation, and the permittivity, which will be discussed later (see Sec.~\ref{Sec:Epsilon}), the bulk packing fraction $\eta$, the ionic strength $I$ of the bulk liquid, the radii
of the three particle types $R_j$, and the parameter $\chi = \frac{9\pi l_\mathrm{B}}{R_1}$ (see Appendix \ref{Sec:ApE}) have to be fixed. When choosing these parameters, we took the respective values for water as guidance,
resulting in particle radii $R_1=R_2=R_3=\SI{1,5}{\angstrom}$ and $\chi  \approx \SI{1.05e4}{}$, where the vacuum Bjerrum length is $l_B = \SI{56.8}{\nano\meter}$.
Furthermore, inspired by Ref. \cite{Roth2002}, the bulk packing fraction is set to $\eta=0.4257$, which together with the chosen particle diameter corresponds to the bulk
number density $\rho_{\text{tot},\mathrm{b}}=\rho_{1,\mathrm{b}}+\rho_{2,\mathrm{b}}+\rho_{3,\mathrm{b}}=\SI{50}{\Molar}\approx\SI{30}{\per\cubic\nm}$.
Finally, the ionic strength is set to $I=\SI{100}{\milli\Molar}\approx\SI{6e-2}{\per\cubic\nm}$.

\section{Results and Discussion\label{Sec:ResultsDiscussion}}
\subsection{Structure of the permittivity\label{Sec:Epsilon}}
\begin{figure}
	\includegraphics{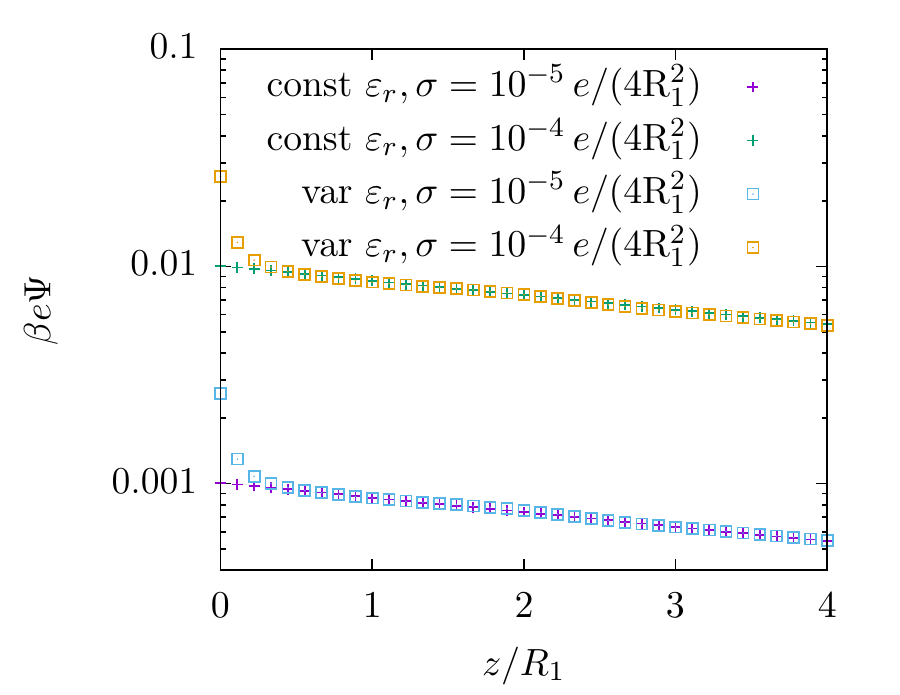}
	\caption{Electrostatic potential $\Psi(\vec{r})$ as function of the distance $z$ from the wall for homogeneously charged walls, i.e., $\sigma=\text{const}$. Due to the homogeneous charge distribution $\Psi$ does not depend on $x$ and $y$. The curves differ both in the strength of the wall charge ($\sigma=\SI{e-4}{\elementarycharge\per(4\square R_1)}$ and $\sigma=\SI{e-5}{\elementarycharge\per(4\square R_1)}$) and in the way the permittivity is incorporated. For the crosses the permittivity in the whole system is constant ($\epsilon_r = 80$) and for the squares the permittivity at each position is calculated according to Eq. \eqref{eq:perm}. Albeit being vastly different, both treatments of the permittivity produce nearly identical results. All differences occur on the length scale of fractions of the particle radius, where the varying permittivity leads to stronger potentials at the wall.\label{fig:perm}}
\end{figure}

Before moving on to the discussion of the various charge patterns in the following sections \ref{Sec:Constant} - \ref{Sec:VariousPatterns}, this paragraph focuses on the structure of the permittivity used in the present study.
It has previously been shown \cite{Kaatze1984}, that a solvent density dependent, linear interpolation of the permittivity between its vacuum value $\epsilon_r^{(0)}=1$ and its value for the pure system (water) $\epsilon_r^{(1)}=80$
matches the behavior of fluid mixtures very well. We have therefore compared two different approaches to set the permittivity in the current study: one is a constant permittivity $\epsilon_r=80$ throughout the
whole system, the other one is a linear interpolation between the two extreme values, scaled with the weighted solvent density:
\begin{equation}
	\epsilon_r(\vec{r}) = \epsilon_r^{(0)} + \frac{3(\epsilon_r^{(1)}-\epsilon_r^{(0)})}{4\pi\rho_{1,\mathrm{b}}R_1^3}\Int{\mathcal{V}}{3}{r'}\rho_1(\vec{r}-\vec{r'})\omega^{(3)}_1(\vec{r});
	\label{eq:perm}
\end{equation}
the structure of this equation is adopted from the FMT approach. In Eq. \eqref{eq:perm} the solvent density is essentially averaged over a particle radius, whereas for the constant
permittivity the solvent density is basically averaged over an infinite region. Calculating the density profile for a homogeneous charge distribution $\sigma(\vec{s})=\text{const}$,
we have compared these two approaches for the permittivity. In Fig. \ref{fig:perm}, the electrostatic potential
$\Psi$ for two homogeneous wall charges $\sigma=\SI{e-4}{\elementarycharge\per(4\square R_1)}$ and $\sigma=\SI{e-5}{\elementarycharge\per(4\square R_1)}$ and for the two cases of the permittivity treatment is shown as
function of the normal distance $z$ from the wall. Since there is no lateral variation of the surface charge distribution, there is consequently also no dependence of the electrostatic potential on the lateral position, which is
why only the  $z-$dependence is shown here. Although the two expressions used for the permittivity are quite distinct, the two approaches yield rather similar results. All differences solely occur on a length scale of
fractions of the particle radius $R_1$ away from the wall, where the varying permittivity $\epsilon_r$ leads to stronger potentials. This is due to the vanishing solvent density $\rho_1(\vec{r})=0$ for distances smaller than a
particle radius $z<R_1$, caused by steric repulsion. Therefore, close to the wall the permittivity decreases, which in turn causes an increasing potential $\Psi$. However, this increase is restricted to the close proximity of the
wall, leading to basically undistinguishable profiles even at distances of the order of the particle radius. Since treating the permittivity according to Eq. \eqref{eq:perm} is computationally very
costly and the benefits are apparently minor, we treat the permittivity as $\epsilon_r(\vec{r}) = \text{constant} = 80$ throughout the whole system, i.e., for all $\vec{r}\in\mathcal{V}$ with $\mathcal{V} = \{\vec{r}\in\mathbb{R}^3|(x,y,z>0) = (\vec{s},z>0)\}$.

\subsection{Constant wall charge distribution \label{Sec:Constant}}
\begin{figure}[!h]
	\includegraphics{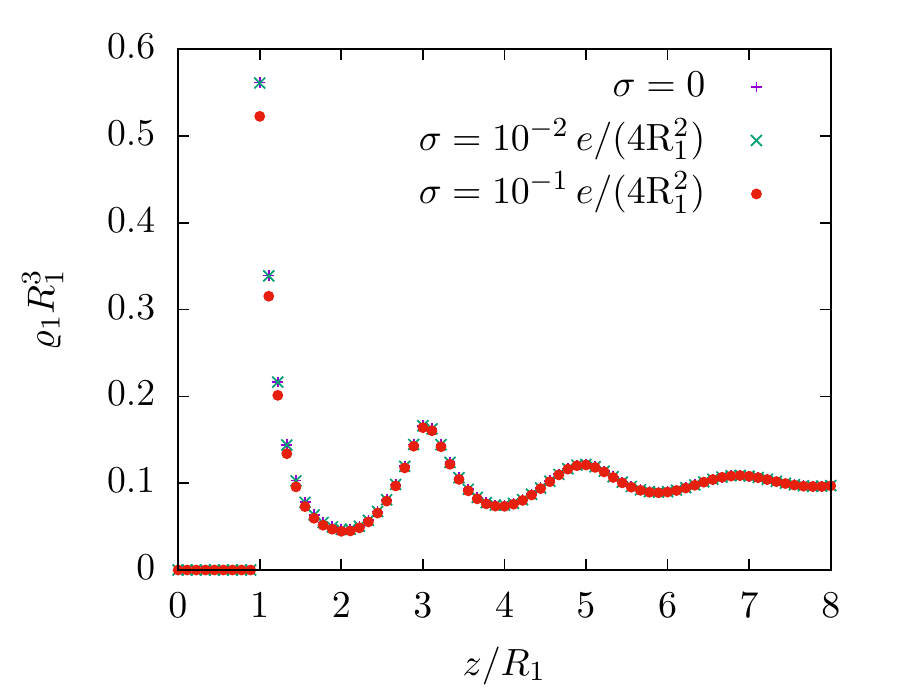}
	\caption{Solvent density $\rho_1(\vec{r})$ as function of the distance $z$ from the wall for three cases of homogeneous wall charges. The case $\sigma = 0$ corresponds to an uncharged wall. The resulting profile is caused only by steric repulsion of the hard spheres. One can see, that only for high wall charges, i.e., $\sigma=\SI{e-1}{\elementarycharge\per(4\square R_1)}\approx\SI{16}{\micro\coulomb\per\centi\meter\squared}$, the density profile deviates from the purely hard sphere profile. For these wall charges the density of the solvent decreases close to the wall, however the amplitude and the range of these changes are rather small. Still, the layer structure, caused by the hard spheres, is predominant.\label{fig:constSolvent}}
\end{figure}

After studying the influence of the permittivity and describing the form used in the present study in the previous Sec.~\ref{Sec:Epsilon}, we now turn towards the analysis of various surface charge patterns,
where we first focus on the simplest case of a spatially constant surface charge distribution $\sigma=\text{const}$ and vary solely its strength. First, we studied the case
of a vanishing surface charge $\sigma=0$. In this case, due to the hard sphere nature of the model, close to the wall layering of the particles  occurs (see Fig. \ref{fig:constSolvent}).
This is an expected result, as it has been reported in numerous previous studies (e.g., Ref. \cite{Roth2002}). If the strength of the surface charge density is slowly increased, one can identify the
effects introduced in the system via electrostatics. In Fig. \ref{fig:constSolvent} the solvent densities for three different surface charge strengths is shown, ranging from a neutral wall ($\sigma = 0$) to a
highly charged wall with $\sigma=\SI{e-1}{\elementarycharge\per(4\square R_1)}\approx\SI{16}{\micro\coulomb\per\centi\meter\squared}$. As can be seen in Fig. \ref{fig:constSolvent}, only for high wall
charges, the density profiles of the solvent start to deviate from the ones found for pure hard spheres with no electrostatic addition. For these high wall charges the solvent density decreases in close proximity to
the charged substrate. However, the size and the range of these deviations are relatively small and even for distances of about two particle radii away from the wall, the profiles reduce to the purely hard
sphere ones.

\begin{figure}[!h]
	\includegraphics{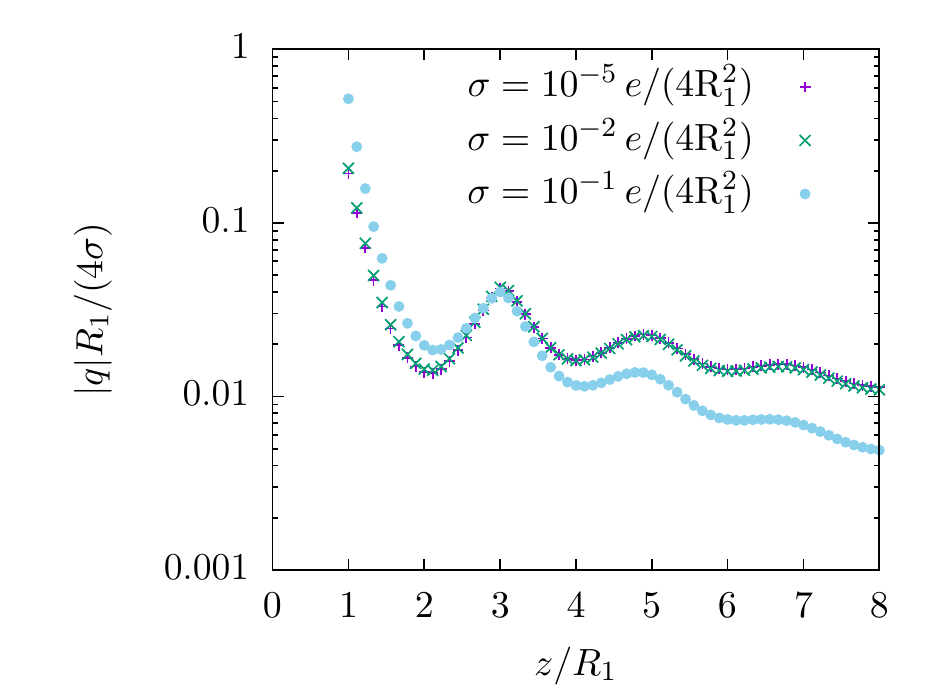}
	\caption{Reduced charge density $q(\vec{r}) = e(\rho_2(\vec{r}) - \rho_3(\vec{r}))$ as function of the distance $z$ to the wall for three cases of homogeneous wall charges. For all three cases one can clearly see the layer structure in the densities, which is caused by the hard sphere nature of the particles. Similar to the solvent density, the reduced charge density stays more or less constant for a wide range of wall charge strengths and deviates noticeably from the low wall charge behavior only for high wall charges, i.e., $\sigma=\SI{e-1}{\elementarycharge\per(4\square R_1)}\approx\SI{16}{\micro\coulomb\per\centi\meter\squared}$. In this case the charge density increases upon approaching closely the wall and decreases for distances larger than three particle radii, when compared with the profiles for smaller wall charges. Also, the hard sphere nature of the particles is still important for high wall charges, as expressed via the layering, which is still apparent.\label{fig:constCharge}}
\end{figure}

By studying the charge densities $q$, and by that the profiles of the ions, a similar observation can be made. In Fig. \ref{fig:constCharge}, the charge density $q$ is shown for a wide range of wall charges $\sigma$, for which
the charge density and thus the profiles of the ions vary only by a proportionality factor. However, for high wall charges, the behavior changes. For these instances, the charge density increases directly at
the wall and decays faster with the distance $z$ from the wall than in the case of lower charge densities, as can clearly be seen in Fig. \ref{fig:constCharge}. In combination with the findings for the solvent
particles, the reason for this effect is obvious. For small wall charges $\sigma$, the fluid reacts by simply swapping co- and counterions so that the total density as well as the solvent density is largely unaffected by this
replacement of ions. However, if the surface charge is becoming too large, this simple replacement is insufficient to neutralize the charge apparent at the wall. Thus, the density of the counterions has to increase even
further by superseding in parts the solvent particles. This explains the decrease in the solvent density $\rho_1$ close to the wall in the case of high wall charges and the change in the shape of the charge density profiles.

\begin{figure}
	\includegraphics{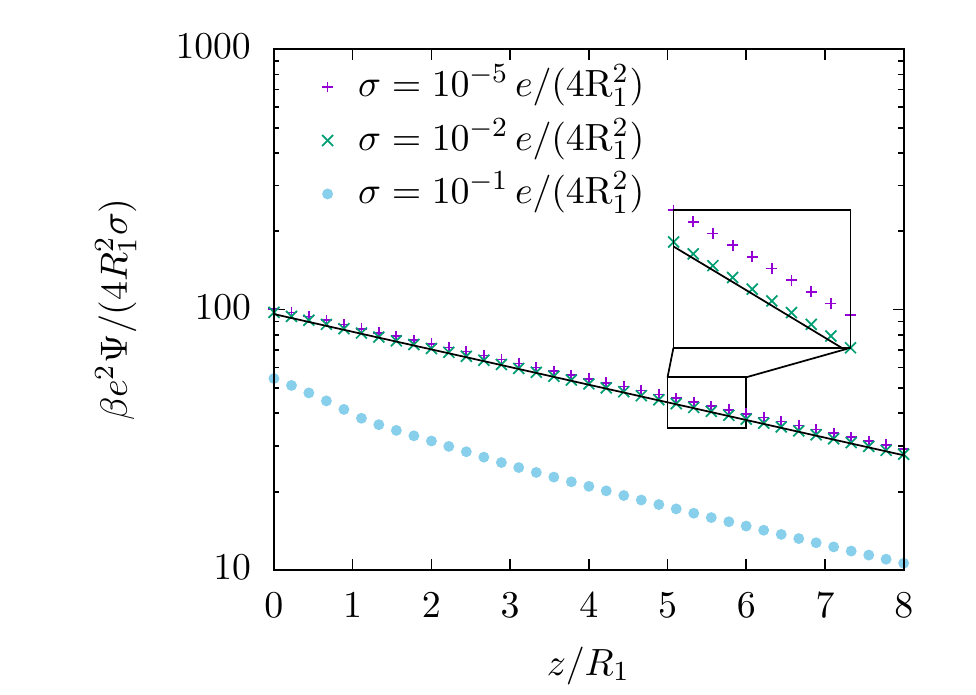}
	\caption{Reduced electrostatic potential $\beta e \Psi$ as function of the distance $z$ from the wall for three cases of homogeneous wall charges (various symbols). For all three cases one can clearly see the linear behavior in the region $z<R_1$, because there is no charge present. Also, in all three cases the potential decreases monotonously with increasing distance $z$. However, in the case of high wall charge densities, i.e., $\sigma=\SI{e-1}{\elementarycharge\per(4\square R_1)}\approx\SI{16}{\micro\coulomb\per\centi\meter\squared}$ (blue circles), there is an apparent change in the behavior: the amplitude of the potential right at the wall decreases, and the profile decays faster in normal direction. Additionally, the results for the wall charge density $\sigma=\SI{e-2}{\elementarycharge\per(4\square R_1)}$, calculated within the model used in Ref. \cite{Mussotter2018}, is shown here as a black line (see the inset). This prediction matches the present results remarkably well. \label{fig:constPots}}
\end{figure}

In addition to the number density profiles, we also studied the profile of the electrostatic potential $\beta e \Psi$ occurring in the case of charged substrates.
In Fig. \ref{fig:constPots} it is shown as function of the distance $z$ from the wall for three strengths of the wall charge density. The observations here are similar
to the previously discussed cases of the number density profiles. For most of the range of wall charges $\sigma$ studied here, the potential is only varying by a proportionality factor
so that the overall shape of the decay of the potential with increasing distances $z$ from the wall stays the same as the number density profiles. Again, this behavior changes for high values of the surface charge density $\sigma$.
As previously seen for the particle profiles in Figs. \ref{fig:constSolvent} and \ref{fig:constCharge}, the decay of the electrostatic potential $\Psi$ changes upon approaching the wall charge density
$\sigma=\SI{e-1}{\elementarycharge\per(4\square R_1)}\approx\SI{16}{\micro\coulomb\per\centi\meter\squared}$. For sufficiently high wall charges, the potential value right at the wall drops below the corresponding value
for lower wall charges, and the decay converges more slowly to the asymptotic behavior with the Debye length as decay length.
This change of behavior can be explained in terms of the observations made for the charge distribution $q$, as a higher absolute value of the charge (the sign
is naturally the opposite of the sign of the wall charge) leads to a stronger screening of the wall charges and thus to smaller values of the electrostatic potential $\Psi$. It also clearly marks the range
of surface charge strengths, within which the linear approximation breaks down. 
Furthermore, Fig. \ref{fig:constPots} shows the result for the corresponding situation of a homogeneous wall charge $\sigma = \SI{e-2}{\elementarycharge\per(4\square R_1)}$ calculated within the framework
of Ref. \cite{Mussotter2018} (black line, see the inset). This result is a perfect match of our findings for the wall charge within the linear regime. This is remarkable, as the framework of Ref. \cite{Mussotter2018} uses a heavily 
simplified fluid model. Still, as we shall see in the following sections \ref{Sec:Sinus} and \ref{Sec:VariousPatterns}, it appears to provide reliable results for the electrostatic potential $\Psi$.

With these insights into the general ranges for which the surface charge $\sigma$ leads to structural effects even for a homogeneously charged wall, we move on towards more complex surface charge distributions.

\subsection{Sinusoidal wall charge \label{Sec:Sinus}}
\begin{figure}
	\includegraphics{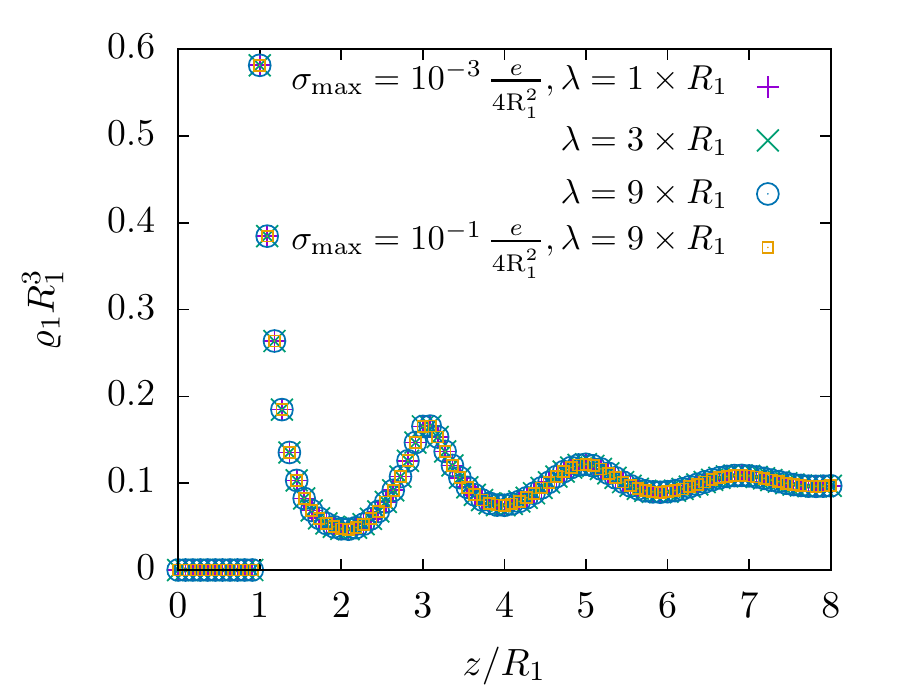}
	\caption{Solvent density $\rho_1(\vec{r})$ for a sinusoidal charge distribution as function of the distance $z$ from the wall for two values of the wall charge strength [$\sigma_{\text{max}} =$ \SIlist{e-3;e-1}{\elementarycharge\per(4\square R_1)}] and three values of the period length [$\lambda = 1\times\mathrm{R}_1,3\times\mathrm{R}_1,\text{ and }9\times\mathrm{R}_1$]. The two wall charge amplitudes are taken from the two regimes found in Sec.~\ref{Sec:Constant}, with $\sigma_{\text{max}} = \SI{e-3}{\elementarycharge\per(4\square R_1)}$ being an exemplary value for the linear response regime. The other value of the wall charge amplitude, $\sigma_{\text{max}} = \SI{e-1}{\elementarycharge\per(4\square R_1)}$, is taken from the suspected non-linear response regime, as identified in Sec.~\ref{Sec:Constant}. There is no visible lateral variation of the profiles, because the laterally varying surface charge density is not strong enough to influence the neutral solvent number density profiles. The profiles show the well-known layering of hard spheres close to a hard wall. When compared with Fig. \ref{fig:constSolvent}, it is clearly visible, that neither the wall charge amplitude nor the period length of the wall charge distribution $\sigma(\vec{s})$ has any influence on the structure of the solvent density $\rho_1(\vec{r})$.\label{fig:sinSolvent}}
\end{figure}

As a first step towards more complex surface charge patterns, we first turn towards a one-dimensional sinusoidal charge distribution 
\begin{equation}
	\sigma(x,y)=\sigma_{\text{max}}\sin(2\pi x/\lambda)
	\label{eq:sinWallcharge}
\end{equation}
with amplitude $\sigma_{\text{max}}$ and wavelength $\lambda$. In contrast to the previous cases studied in Sec.~\ref{Sec:Constant},
there is no net charge on the wall. Due to this absence of a net charge one expects a very short-ranged influence of the wall structure on that of the fluid.
As can be seen in Fig. \ref{fig:sinSolvent}, the effect of the surface charge inhomogeneity on the solvent is indeed limited to a very short range with a small amplitude, as no effects of solvent particle
displacement are visible. This hints at small charge density values inside the fluid. In fact, the difference between the two regions of linear and non-linear fluid response (appearing for $\sigma = \text{const}$) seems to vanish,
or at least the transition seems to be shifted as a function of the wall charge amplitude $\sigma_{\text{max}}$. This is in line with the observation that in the case of the sinusoidal surface charge distribution the solvent density
$\rho_1(\vec{r})$ exhibits no visible deviations from the profiles found for a purely hard sphere system without any electrostatics (see Fig. \ref{fig:sinSolvent}).

\begin{figure}
	\includegraphics{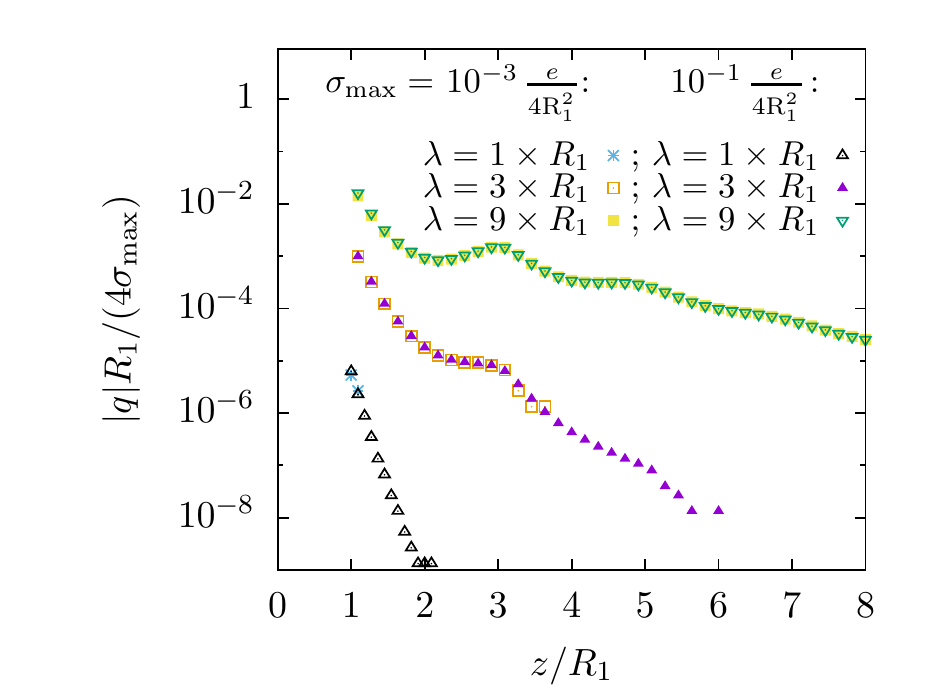}
	\caption{Reduced charge density $q(\vec{r})$ as function of the distance from the wall $z$ evaluated at the maximal amplitude, i.e., at $x=\lambda/4$, for two values of the wall charge strength [$\sigma_{\text{max}} =$ \SIlist{e-3;e-1}{\elementarycharge\per(4\square R_1)}] and three values of the period length [$\lambda = 1\times\mathrm{R}_1,3\times\mathrm{R}_1,\text{ and }9\times\mathrm{R}_1$]. The two wall charge amplitudes are taken from the two regimes found in Sec.~\ref{Sec:Constant}, with $\sigma_{\text{max}} = \SI{e-3}{\elementarycharge\per(4\square R_1)}$ being an exemplary value for the linear response regime. The other value of the wall charge amplitude, $\sigma_{\text{max}} = \SI{e-1}{\elementarycharge\per(4\square R_1)}$, is taken from the suspected non-linear response regime, as identified in Sec.~\ref{Sec:Constant}. The lateral variation of the profiles shown is strictly following the shape of the surface charge distribution, with the lateral position chosen here serving as an example. The known effect of cancellation of significant digits in the number densities generate the loss of precision for the charge density profile, leading to the plateaus visible in the profiles. The wavelength $\lambda$ of the underlying surface charge structure $\sigma(\vec{s})$ (see Eq. \eqref{eq:sinWallcharge}) clearly influences the decay behavior of the charge density. The decay length increases with increasing period length. In order to highlight, that the amplitude $\sigma_{\text{max}}$ of the charge distribution has apparently no further influence than that of a proportionality factor, here the charge density $q$ is reduced accordingly by $\sigma_{\text{max}}$.\label{fig:sinCharge}}
\end{figure}

However, when moving on to the charge density distribution $q(\vec{r}) \equiv e(\rho_2(\vec{r}) - \rho_3(\vec{r}))$, which is shown in Fig. \ref{fig:sinCharge}, there are clear effects specific to the sinusoidal
charge distribution. Similar to Fig. \ref{fig:constCharge}, in Fig. \ref{fig:sinCharge} the charge density distribution $q(\vec{r})$ is shown as a function of the distance $z$ from 
the wall. First, as already suspected from the solvent density profiles $\rho_1(\vec{r})$ in Fig. \ref{fig:sinSolvent}, one can see, that the amplitude of the charge density distribution inside
the fluid is indeed significantly smaller than in the case of a electrostatically homogeneous wall. This small amplitude leads to apparent "plateaus" of the profiles, which occur for numerical reasons.
This effect, however, is of no importance for any of the following conclusions. Second, the profiles
in Fig. \ref{fig:sinCharge} clearly show, that the wavelength $\lambda$ of the surface charge pattern has a strong influence on the decay behavior of the charge density inside the fluid.
With increasing periodicity, i.e., increasing $\lambda$, the decay length of the charge density profiles increases, too. Whereas the charge density decays to (numerically) vanishing values within a length of only a few particle
radii for a wavelength of $\lambda = 1\times\mathrm{R}_1$, it seems to converge to an asymptotically exponential decay for larger wavelengths of the surface charge pattern $\sigma(\vec{s})$. Furthermore, as
stated previously for the solvent density profiles $\rho_1(\vec{r})$ in Fig. \ref{fig:sinSolvent}, the profiles for the charge density $q(\vec{r})$ in the fluid show no dependence on the amplitude
of the surface charge distribution, although the two values chosen for the amplitude $\sigma_{\text{max}}$ were taken from the two regimes of different fluid reaction, which were found in Sec.~\ref{Sec:Constant}.

\begin{figure}[h!]
	\includegraphics{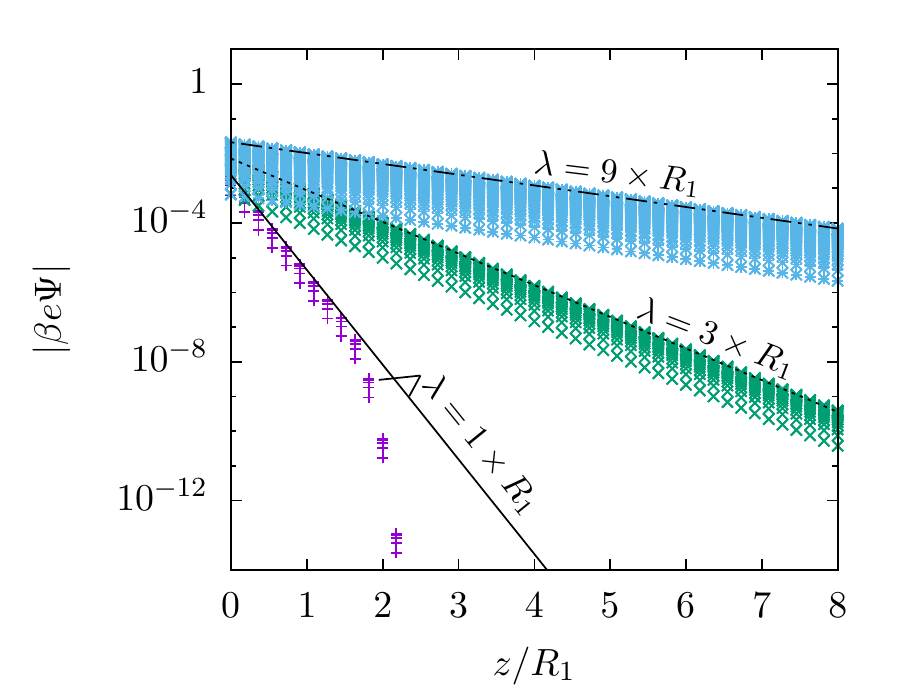}
	\caption{Electrostatic potential $\Psi$ as a function of the distance $z$ from the wall for all lateral positions $(x,y)$ studied for three period lengths ($\lambda = 1\times\mathrm{R}_1,3\times\mathrm{R}_1,\text{ and }9\times\mathrm{R}_1$) of the sinusoidal wall charge distribution (see Eq. \eqref{eq:sinWallcharge}). The amplitude of the wall charge is set to $\sigma_{\text{max}} = \SI{e-3}{\elementarycharge\per(4\square R_1)}$ for all these period lengths. Different data points of one color (shaded areas) for one normal distance correspond to various lateral positions $(x,y)$ along one period of the surface charge pattern. The laterally varying strength of the surface charge leads to these broad ranges of data points. The width of these ranges provides information about the strength of the lateral variation: the wider the range, the stronger is the lateral variation of the electrostatic potential. The straight lines correspond to an exponential decay with a decay length, which results from a combination of the Debye length and the corresponding inverse dominant length scale following the prediction in a previous study (see Eq. \eqref{eq:sinProp} and Ref. \cite{Mussotter2018}). The agreement between this prediction and the present data is remarkable.\label{fig:sinPots}}
\end{figure}

Finally, as for the homogeneous wall charge (see Sec.~\ref{Sec:Constant}), we investigated the electrostatic potential $\Psi$ for the case of a sinusoidal surface charge pattern $\sigma(\vec{s})$
(see Eq. \eqref{eq:sinWallcharge}). In Fig. \ref{fig:sinPots} the electrostatic potential $\Psi$ is shown for three different wavelengths of the surface charge ($\lambda = 1\times\mathrm{R}_1,3\times\mathrm{R}_1,\text{ and } 9\times\mathrm{R}_1$).
Due to the lack of an effect, as has been shown in Figs. \ref{fig:sinSolvent} and \ref{fig:sinCharge}, here the amplitude of the surface charge pattern is kept constant
at $\sigma_{\text{max}} = \SI{e-3}{\elementarycharge\per(4\square R_1)}$. The various data points for a single value of $z$ correspond to different lateral positions $x$ along one
period of the surface charge pattern. First, although the different wavelengths clearly influence the charge density $q$ of the fluid right next to wall, as can be inferred from 
Fig. \ref{fig:sinCharge}, there seems to be no significant effect of the surface charge period length on the strength of the electrostatic potential $\Psi$ at the wall.
Second, also like seen for the charge density $q$, the electrostatic potential clearly decays exponentially with increasing distance $z$ from the wall.
The decay length of $\Psi$ also clearly depends on the wavelength of the surface charge, where an increasing wavelength $\lambda$ leads to an increased decay length. Note, that the deviation from the
exponential decay for the case $\lambda=1\times\mathrm{R}_1$ is due to the numerically caused lack of precision in determining the charge density (see Fig. \ref{fig:sinCharge}).
In Fig. \ref{fig:sinPots}, in addition to the data points, there are three lines indicating the exponential decay corresponding to the prediction in Ref. \cite{Mussotter2018}. There
the decay as a function of $z$ turned out to be proportional to
\begin{equation}
	\beta e \Psi \propto \exp\left(-\sqrt{\kappa^2 + |\vec{q}^2_{\parallel}|}z\right),
	\label{eq:sinProp}
\end{equation}
where $\kappa$ is the inverse Debye length and $|\vec{q}_{\parallel}|$ is the absolute value of the Fourier component of the dominating lateral pattern of the surface charge distribution, which in the present case of the
sinusoidal surface charge pattern (see Eq. \eqref{eq:sinWallcharge}) equals $2\pi/\lambda$. Note that neither the amplitudes of the lines 
shown in Fig. \ref{fig:sinPots}, nor the decay lengths are fitting parameters. The lines strictly follow the results obtained from the counterpiece of Eq. \eqref{eq:sinProp} in Ref. \cite{Mussotter2018}. Although the system
described in Ref. \cite{Mussotter2018} was more basic and the description of the fluid was rather simplistic, the findings deliver remarkably accurate predictions when compared with the results of the present analysis using a
significantly more elaborate fluid description. With this very good agreement with previous, more simplistic approaches, we move on to compare further, more complex, surface charge distributions $\sigma(\vec{s})$ with the
predictions made in Ref. \cite{Mussotter2018}.

\subsection{Various surface charge patterns\label{Sec:VariousPatterns}}
\begin{figure}
	\includegraphics{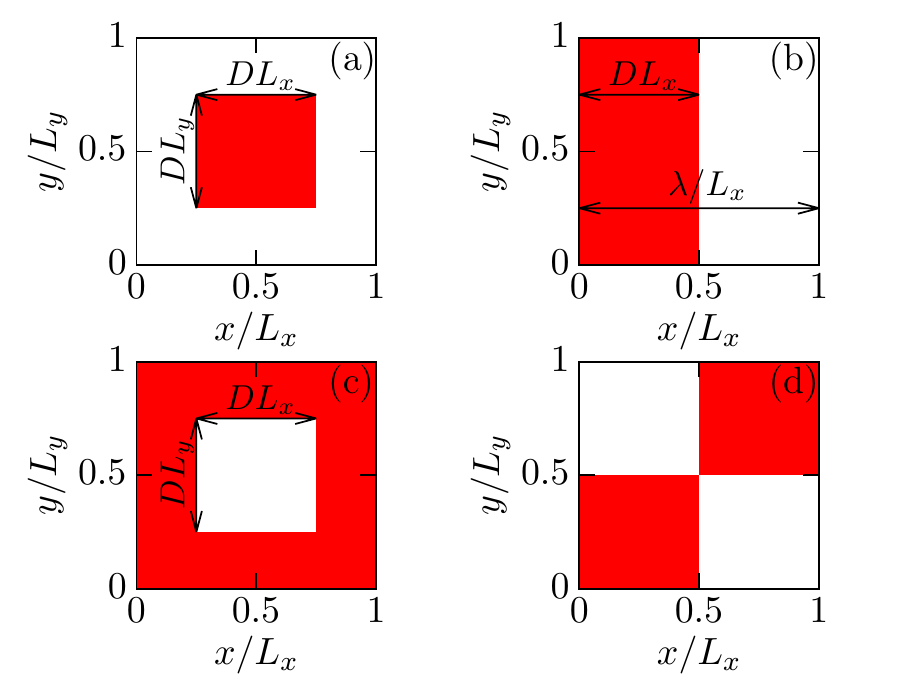}
	\caption{Further surface charge patterns $\sigma(\vec{s})$ considered in the present section. The red areas are regions with non-vanishing surface charges $\sigma(\vec{s}) = \sigma_{\text{max}}$, whereas the white regions are uncharged. All these patterns correspond to the elementary cell of the surface charge pattern, which is periodically continued in both lateral directions along the wall surface. In panel (a), this leads to a pattern of individual squares, in panel (b), this leads to a striped pattern, in panel (c), this leads to a pattern with holes, and in the case of panel (d), one ends up qith a checkerboard pattern.\label{fig:chargePatterns}}
\end{figure}

Moving on from the somewhat simple surface charge patterns $\sigma(\vec{s})$ discussed in Secs. \ref{Sec:Constant} and \ref{Sec:Sinus}, in the present section we investigate more complex
charge distributions. The four principal cases of charge patterns studied here are shown in Fig. \ref{fig:chargePatterns}, where both lateral lengths $L_x$ and $L_y$, the dimensionless charge width $D$, and the amplitude
$\sigma_{\text{max}}$ have been varied throughout the calculations.

\begin{figure}
	\includegraphics{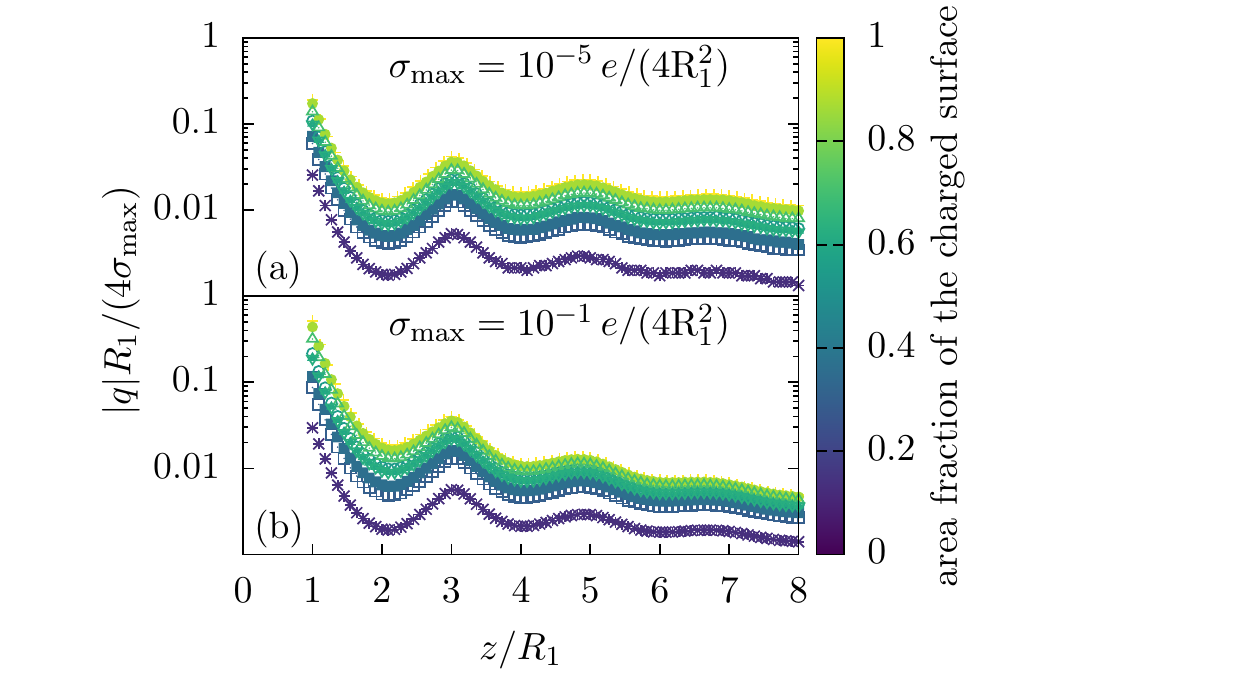}
	\caption{Reduced charge density $q(\vec{r})$ as function of the distance $z$ from the wall and of the area fraction of the charged area per unit cell area for two cases of the wall charge strength: $\sigma_{\text{max}} =$ \SI{e-1}{\elementarycharge\per(4\square R_1)} (panel (b)) and $\sigma_{\text{max}} =$ \SI{e-5}{\elementarycharge\per(4\square R_1)} (panel (a)). Since the lateral variation of the electrostatic potential $\Psi$ de facto disappears for $z\gtrsim R_1$ for all the situations shown here, there is no visible dependence of the displayed profiles on the lateral position. All the profiles shown here correspond to various realizations of the charge patterns shown in Fig. \ref{fig:chargePatterns}. It turns out that the actual configuration (i.e., Figs. \ref{fig:chargePatterns}(a), (b), (c), or (d)) is not important. Instead only the averaged surface charge, given by the area fraction of the charged surface, appears to matter. One can clearly see, how the increase of the averaged wall charge also leads to increased charge densities inside the fluid and how the curves for increasing area fraction converge towards the ones for the homogeneous wall as shown in Fig. \ref{fig:constCharge}. Furthermore, the range of averaged surface charges shown in panel (b) coincides with the transition region as identified in Sec.~\ref{Sec:Constant}. Therefore, panel (b) resolves this region in more detail.\label{fig:allCharge}}
\end{figure}

This variation of parameters leads, inter alia, to a variation of the effective surface charges, i.e., the averaged surface charge strengths $\sigma_{\text{av}}$, as the area fraction of the charged area compared to the total area of the surface charge unit cell $L_x\times L_y$ is changed.
The influence of this change can be seen in Fig. \ref{fig:allCharge}. Here, the charge density $q(\vec{r})$ is shown as function of the distance $z$ from the wall. Note, that for the cases shown here,
which all correspond to the case of the shortest lateral wavelength considered ($\lambda = 1\times \mathrm{R}_1$),
the dependence of the potential $\Psi$ on the lateral position $\vec{s}=(x,y)$ disappears for $z\gtrsim R_1$. Therefore, the shown charge densities do not exhibit any visible lateral dependence. In Fig. \ref{fig:allCharge} one can clearly
see that, independent of the surface charge amplitude $\sigma_{\text{max}}$, all surface charge patterns lead to qualitatively similar results. The charge decays exponentially with $z$ and shows clear signs of the hard core nature
of the particles at small distances from the wall. Furthermore, one can infer from these graphs, that the charge density profiles $q(z)$ smoothly converge towards the ones found in Fig. \ref{fig:constCharge} as the area
fraction of the charged surface is increased. Especially in the case of the stronger wall charge amplitude ($\sigma_{\text{max}} = \SI{e-1}{\elementarycharge\per(4\square R_1)}$, Fig. \ref{fig:allCharge}(b)) this is interesting, because these profiles
and their associated range of averaged surface charges coincide with the transition region from a linear to a non-linear fluid reaction, as has been found previously (Sec.~\ref{Sec:Constant}). The profiles in Fig. \ref{fig:allCharge} corresponding to
a small area fraction and a weak averaged surface charge, respectively, still exhibit a linear response behavior, because the reduced profiles for both wall charge amplitudes
$\sigma_{\text{max}} =$ \SIlist{e-1;e-5}{\elementarycharge\per(4\square R_1)} are the same. Thus the amplitude is solely a proportionality factor, which matches the behavior characterizing the linear response regime.
However, if the area fraction and thus the averaged wall charge is increased, deviations from the respective profiles for the above two wall
charge amplitudes increase, until finally the two profiles for the fully charged wall match the previous results in Fig. \ref{fig:constCharge}. Therefore, the transition between these regions is smooth and does not show any
sign of a step-like variation.

\begin{figure}
		\includegraphics{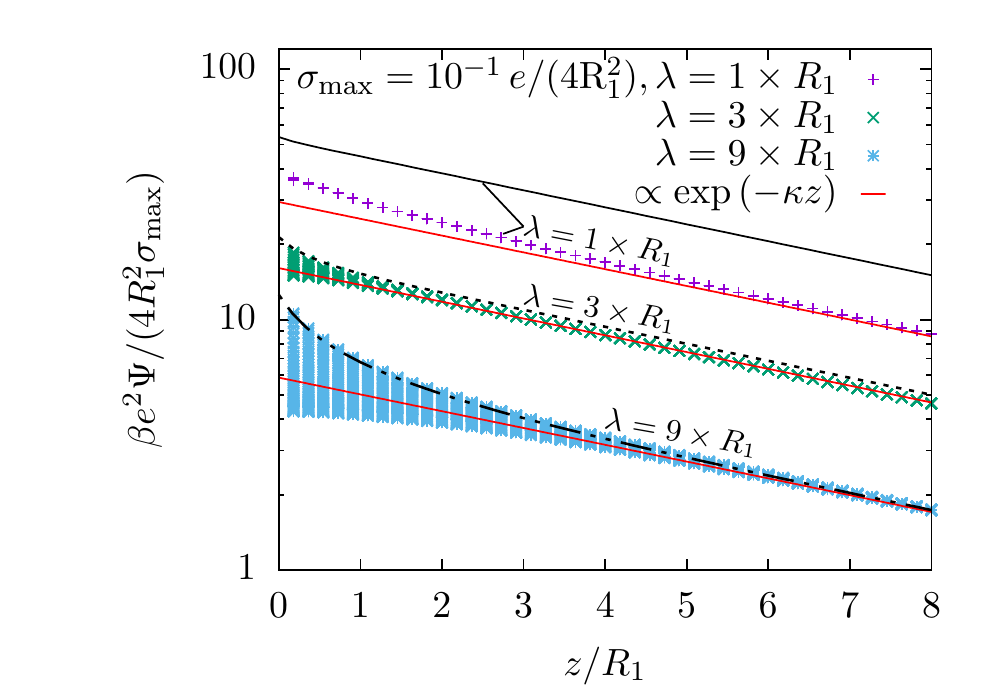}
		\caption{Scaled electrostatic potential $\Psi$ as a function of the distance $z$ from the wall for all lateral positions $\vec{s}$ studied for three wavelengths $\lambda = 1\times\mathrm{R}_1,3\times\mathrm{R}_1\text{, and }9\times\mathrm{R}_1$ of the wall charge distribution; here it is a pulse wave-like charge distribution (see Fig. \ref{fig:chargePatterns}(b) and Eq. \eqref{eq:case3Wallcharge}). The duty cycle $D$, i.e., the dimensionless width of the charged stripe, remains the same ($DL_x=0.5\times\mathrm{R}_1$) for all three cases. Similar to Fig. \ref{fig:sinPots}, the region covered by the spread of the data is due to different lateral positions being shown for the same distance from the wall. Again, this gives information about the strength and the range of lateral variations of the electrostatic potential. However, in contrast to the potentials shown in Fig. \ref{fig:sinPots}, here the wall carries a net charge. This is the reason for the visible long-ranged decay, which corresponds to the exponential decay of the associated net charge with a decay length equal to the corresponding Debye length $\kappa^{-1}$ (see Eq. \eqref{eq:PotAssy}).
		This decay behavior far from the wall occurs for all three cases. However, close to the wall there is a much more complicated decay behavior, which strongly depends on the wavelength of the surface charge pattern. As in previous graphs, the black lines correspond to the results for the same calculations within the framework of Ref. \cite{Mussotter2018}. For the two longer wavelengths, $\lambda = 3\times\mathrm{R}_1\text{ and }9\times\mathrm{R}_1$, these results again match the present ones very well. The deviation occurring for the shortest wavelength is due to the fact, that this case is outside of the linear response regime (see Sec.~\ref{Sec:Constant}). \label{fig:allPots}}
	\end{figure}

Moving on to the case of longer lateral wavelengths, i.e., $\lambda = 3\times\mathrm{R}_1$ and $\lambda = 9\times\mathrm{R}_1$, close to the wall the charge density $q(z)$ starts to exhibit a lateral structure. In contrast to the
profiles shown in Fig. \ref{fig:allCharge}, the lateral position $\vec{s}$ influences the local charge density $q(\vec{s},z)$, for boundary conditions with these longer lateral wavelengths. This lateral variation is
even more visible in the profiles of the electrostatic potential. Thus, in the following the behavior of the electrostatic potential $\Psi$ for these more complicated charge distributions $\sigma(\vec{s})$ is studied, where
we focus on surface charge patterns of the form shown in Fig. \ref{fig:chargePatterns}(b):
\begin{equation}
	\sigma(\vec{s}) = \left\{\begin{array}{ll}
		\sigma_{\text{max}}, & \text{for } N\lambda\leq x\leq (N+D)\lambda\\
			0, &\text{otherwise},
	\end{array}\right.
		\label{eq:case3Wallcharge}
\end{equation}
	with $\sigma_{\text{max}}$ as the amplitude, $N\in\mathbb{Z}$, $\lambda$ as the wavelength, and $D$ being the dimensionless width of the charged stripe. We studied the cases
	$\lambda = L_x = 1\times\mathrm{R}_1,3\times\mathrm{R}_1,\text{ and }9\times\mathrm{R}_1$ with $D\times L_x= 0.5\times\mathrm{R}_1$.
	These choices are taken for the sake of simplicity. The findings discussed in the following can easily be verified also for the other charge distributions shown in Fig. \ref{fig:chargePatterns}. 
	The resulting profiles for the electrostatic potential $\Psi$ are shown in Fig. \ref{fig:allPots}. Here the data are shown together with the asymptotic Debye decay (red solid lines) and with the results for the
	same boundary conditions $\sigma(\vec{s})$, but obtained within the framework of Ref. \cite{Mussotter2018} (black lines). 

	First, we note the offset between the three profiles, which is due to the fact, that the net charge differs for the
	three displayed cases. Here, however, the potential $\Psi$ is reduced with respect to the amplitude $\sigma_{\text{max}}$ only. If one accounts for the different net charges as well by determining the averaged charge and
	reducing $\Psi$ with respect to the averaged surface charge $\sigma_{\text{av}}$ instead of the maximum one $\sigma_{\text{max}}$, all three cases render the
	same asymptotic profile. Second, the wavelength $\lambda$ of the surface charge pattern $\sigma(\vec{s})$ strongly influences the behavior of the potential $\Psi$ close to the wall. With increasing wavelength, the potential
	exhibits a strong dependence on the lateral position, as can be inferred from the range of potential values at $z=0$. Also, the decay length of the electrostatic potential close to the wall strongly increases with
	increasing wavelength $\lambda$. Far away from the wall all three cases clearly match the predictions of an exponential decay with the decay length given by the Debye length $\kappa^{-1}$.
	Finally, the comparison with the results calculated along the lines proposed in Ref. \cite{Mussotter2018} again reveals remarkable agreement, at least for the two larger wavelengths. The results of the calculation within the
	framework of Ref. \cite{Mussotter2018} are obtained in the middle of one of the charged areas. Therefore, they should follow the highest values of the data obtained from the calculations of the present study. This can
	easily be verified.
	The reason for the discrepancy in the case of the smaller wavelength $\lambda=1\times\mathrm{R}_1$ can be found by comparison with the situation discussed in Sec.~\ref{Sec:Constant}. In that section, we found a clear change
	in the behavior of the electrostatic potential $\Psi$ for high wall charges, where the fluid reaction becomes non-linear. In Fig. \ref{fig:allPots}, the effective surface charges of the three cases lie around this transition,
	with the case $\lambda = 9\times\mathrm{R}_1$ still being in the linear regime, the case $\lambda = 1\times\mathrm{R}_1$ being in the non-linear regime, and the case $\lambda = 3\times\mathrm{R}_1$ being very close to the transition.
	Due to that, we find very good agreement for the largest of the wavelengths and increasing deviations for decreasing wavelengths. Especially for the smallest wavelength, $\lambda = 1\times\mathrm{R}_1$, one is clearly in the
	non-linear regime, which indicates the failure of the model used in Ref. \cite{Mussotter2018} (see Fig. \ref{fig:constPots} in Sec.~\ref{Sec:Constant}).

\begin{figure}[h!]
		\includegraphics{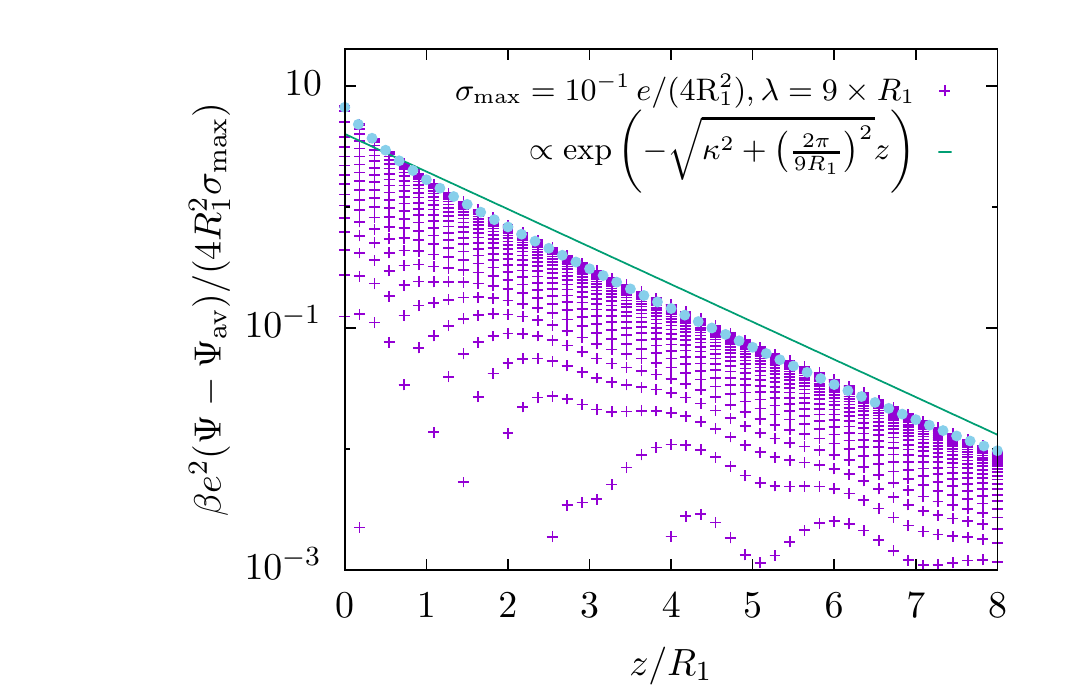}
		\caption{Electrostatic potential $\Psi$ reduced by the surface charge amplitude $\sigma_{\text{max}}$ as a function of the distance $z$ from the wall for all lateral positions $\vec{s}$ studied for the case of an underlying surface charge pattern $\sigma(\vec{s})$ corresponding to Fig. \ref{fig:chargePatterns}(b) with wavelength $\lambda = 9\times\mathrm{R}_1$. Since all lateral positions are shown, there are multiple data points for one distance $z$, leading to the region covered by the spread of the data. Additionally, the asymptotic profile $\Psi_{\text{av}}$ with the Debye length $1/\kappa$ as decay length is subtracted from the data in order to gain insight into the next shorter, subdominant length scale involved (see Eq. \eqref{eq:PotAssy}). The green line corresponds to this shorter length scale as it is obtained from Eq. \eqref{eq:sinProp}. Furthermore, the blue circles depict the results for the same system derived via the framework of Ref. \cite{Mussotter2018}. Not only do the data follow the theoretical predictions very well, also both data sets match remarkably well, despite large differences in the details of the fluid description. \label{fig:potsZoom}}
	\end{figure}

	Finally, we take a closer look at the decay of the electrostatic potential $\Psi$ for the wavelength of $\lambda = 9\times\mathrm{R}_1$. In Fig. \ref{fig:potsZoom}, the asymptotic behavior of the electrostatic potential
	\begin{equation}
		\Psi_{\text{av}}\propto \exp(-\kappa z),\label{eq:PotAssy}
	\end{equation}
	caused by a non-vanishing average charge $\sigma_{\text{av}}$ of the surface with $\kappa$ as the Debye length, is subtracted from the data to study shorter ranged contributions to the behavior of the potential close
	to the wall. As given in Eq. \eqref{eq:sinProp}, the theoretical predictions from Ref. \cite{Mussotter2018} for a
	linear response approximation hints at a decay with a decay length depending on the wavenumber $q_{\parallel}$ of the surface charge pattern $\sigma(\vec{s})$. In fact, there are multiple further exponential
	decays involved, all of which depend on the wavenumber; Eq. \eqref{eq:sinProp} represents only the next smaller (to $1/\kappa$) length scale of the decay of the potential. This prediction is shown as a green solid
	line in Fig. \ref{fig:potsZoom}. Again, we also include the results given by Ref. \cite{Mussotter2018} for the same surface charge distribution (blue circles). The various data points for the same
	distance $z$ correspond to different lateral positions. Close to the wall, the	potential $\Psi$ exhibits a faster decay than the one given by the displayed prediction (green line), which agrees with the expected occurrence
	of further short-ranged decays influencing the behavior in close proximity to the wall. However, within the intermediate range of distances from the wall ($6\times\mathrm{R}_1\gtrsim z\gtrsim 2\times\mathrm{R}_1$), the data
	closely follow the lowest order predictions from Ref. \cite{Mussotter2018}. The present data clearly shows an exponential decay, with the decay indeed given by Eq. \eqref{eq:sinProp}. Nevertheless, even upon	closer
	examination, our findings match with the full corresponding results from Ref. \cite{Mussotter2018} (blue dots) remarkably well.

\begin{figure}[h!]
		\includegraphics{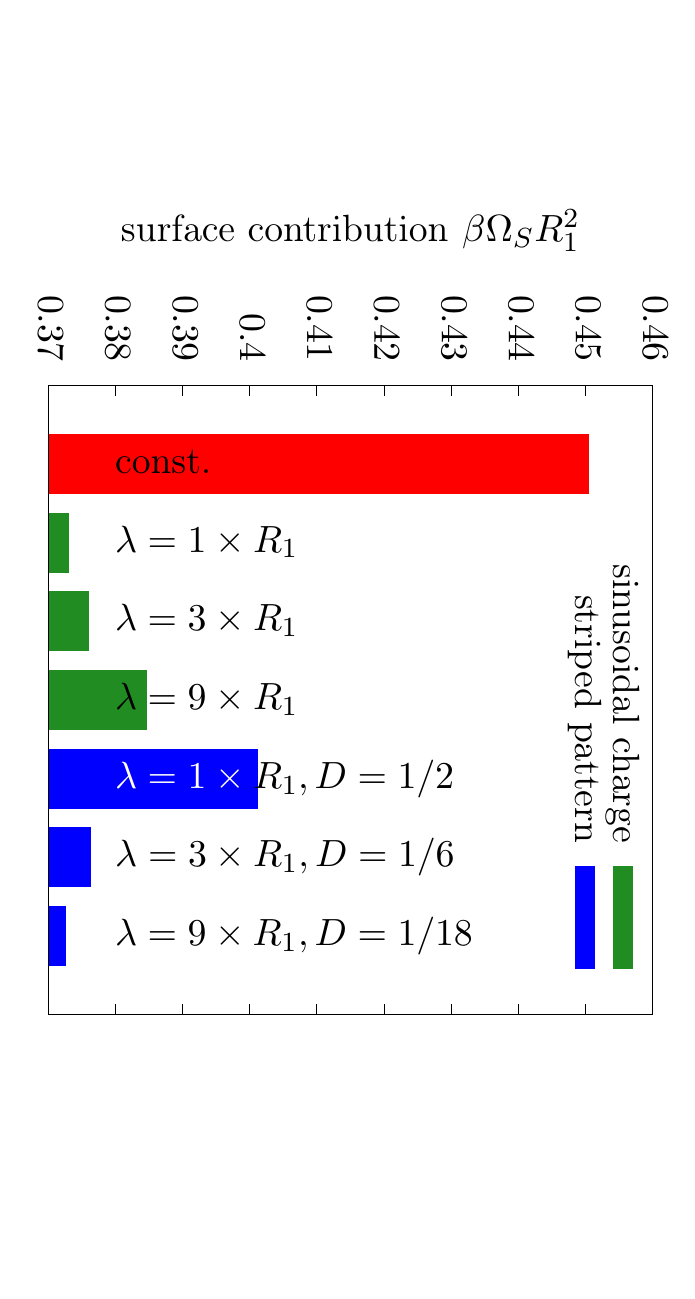}
		\caption{Dimensionless surface contribution $\Omega_S$ to the grand potential for the different surface charge configurations studied here (see Eq. \eqref{eq:Gamma}). For all cases, the surface charge amplitude is set to $\sigma_{\text{max}} = \SI{e-1}{\elementarycharge\per(4\square R_1)}$, or $\sigma = \SI{e-1}{\elementarycharge\per(4\square R_1)}$ for the constant wall charge, respectively. For a purely repulsive hard wall, the surface contribution is $\beta\Omega_S R_1^2 = 0.36972$ (not shown here). For values of the wall charge smaller than the one shown here, the variations of $\Omega_S$ are negligibly small. Additionally, for the striped pattern the duty cycle $D$, i.e., the dimensionless width of the charged stripe, is kept constant ($DL_x=0.5\times\mathrm{R}_1$). As the overall charge in the case of the constant wall charge is the highest by far, this case clearly shows the largest surface contribution $\Omega_S$. However, for the other cases with spatially varying surface charge the surface contribution $\Omega_S$ depends on the wavelength of the pattern and on the average charge of the wall. \label{fig:Surfacetensions}}
	\end{figure}

	Finally, we compare the previously discussed surface charge patterns with respect to the surface contribution (see Refs. \cite{Dietrich1988, Schick1990})
\begin{equation}
	\Omega_S = \frac{\Omega^{\text{eq}} + pV}{A}
	\label{eq:Gamma}
\end{equation}
of the grand potential $\Omega^{\text{eq}}$, where $p$ is the bulk pressure, $V = |\mathcal{V}|$ is the size of the system, and $A = |\partial\mathcal{V}|$ is the area of the charged wall. Note that the quantity $\Omega_S$, which has the dimension of energy per area, is sometimes called "surface tension", whereas some authors decompose it into the surface tension of a uniform wall, the line tension, etc. $\Omega_S A$ measures the cost of free energy to create an area $A$ with the respective surface charge pattern.
	The resulting values for the various configurations of the wall charge are shown in Fig. \ref{fig:Surfacetensions}. There, the surface contributions $\Omega_S$ are shown for a surface charge amplitude of $\sigma_{\text{max}} = \SI{e-1}{\elementarycharge\per(4\square R_1)}$, or $\sigma = \SI{e-1}{\elementarycharge\per(4\square R_1)}$ for the constant wall charge, respectively.
For smaller wall charge amplitudes, there is no significant effect visible. For the shown strength of the surface charge, however, there are some interesting features. First, the surface contribution $\Omega_S$ for the constant surface charge distribution is much higher than for all the other cases. Because the overall charge in this case is the highest, this large influence on the structure is understandable. Second, in the case of the sinusoidal wall charge distribution, the surface contribution $\Omega_S$ clearly increases with the wavelength of the surface structure. This is understandable, too, because the lateral variation of, e.g., the electrostatic potential, and also the range of this variation normal to the surface, becomes more pronounced for increasing wavelengths (see Figs. \ref{fig:sinPots} and \ref{fig:allPots}), which reflects the influence of the surface on the fluid. Thus, the surface contribution $\Omega_S$ increases for larger wavelengths. This effect, however, seems to be reversed if the surface is arranged as shown in Fig. \ref{fig:chargePatterns}(b), i.e., as a striped pattern. For this case Fig. \ref{fig:Surfacetensions} indicates, that the surface contribution $\Omega_S$ decreases for increased wavelengths. However, in contrast to the sinusoidal charge pattern, the striped pattern carries an average charge, which increases with decreasing wavelength ($DL_x$ is kept constant, see Fig. \ref{fig:Surfacetensions}). This competition of increasing range and decreasing average charge leads to the observed behavior. Thus, the surface contribution $\Omega_S$ nicely echoes the previous findings, for which the fluid structure depends on both the average charge of the wall, especially for small scale surface charge patterns (see Fig. \ref{fig:allCharge}), and the wavelength of the surface charge distribution (see Figs. \ref{fig:allPots} and \ref{fig:potsZoom}).

The observed dependence also provides information about the solubility of particles carrying a surface charge. As mentioned above, the surface contribution $\Omega_S A$ measures the cost of free energy to form an interface of area $A$. Therefore, large surface contributions lead to weak solubilities, because creating the interface is energetically costly. The planar surface charge patterns studied here can be regarded as surface segments of particles, which are large compared to the fluid constituents, i.e., for which a planar surface is an acceptable approximation. Hence, we find the solubility of such large particles, which carry a surface charge pattern associated with the corresponding surface contribution $\Omega_S A$ (see Fig. \ref{fig:Surfacetensions}), to vary with the wavelength and the average charge of the pattern.

	\section{Conclusions and summary\label{Sec:ConclusionsSummary}}

	In the present study the effects of surface charge heterogeneities on a nearby electrolyte solution has been investigated with respect to the density profiles of all three fluid components and the electrostatic potential inside 
	the system. The fluid comprises a neutral solvent and a single univalent salt component. They are treated explicitly as hard spheres by means of classical density functional theory within the
	framework of fundamental measure theory, which has been proven to be a powerful approach to study fluid structures in terms of number density profiles \cite{Evans1979, Evans1990, Evans1992, Roth2002}. In order to gain further insight
	into this system, a variety of surface charge patterns has been studied, starting with the case of a homogeneous wall charge distribution (see Sec.~\ref{Sec:Constant}). For such homogeneously charged walls we have naturally
	found no dependence of any of the profiles on the lateral position. But for increasing distances from the wall (see Figs. \ref{fig:constSolvent}, \ref{fig:constCharge}, and \ref{fig:constPots}) we have been able to identify an exponential decay on the length scale of the Debye length $\kappa^{-1}$ for all studied values of the constant surface charge density.
	Beyond that, the density profiles of the three fluid components are dominated by well-known layering effects caused by the hard sphere nature of all particles.
	Furthermore, for various wall charge strengths we have identified two regimes of the fluid response. For low surface charges we have found a linear response of the fluid, whereas replacement of solvent particles by counterions leads
	to non-linear response phenomena for high surface charge strengths.

	In Sec.~\ref{Sec:Sinus}, replacing the homogeneously charged (and therefore overall charged) wall by a sinusoidal charge distribution (with no overall charge), we have found a strong dependence of both the solute
	densities and the electrostatic potential on the wavelength of the underlying surface charge pattern (see Figs. \ref{fig:sinCharge} and \ref{fig:sinPots}). However, the solvent densities remain de facto unchanged
	upon a change of the wavelength (see Fig. \ref{fig:sinSolvent}). Also, for all studied values, there are no dependences on the amplitude of the surface charge other than a proportionality factor.

	Finally, in Sec.~\ref{Sec:VariousPatterns} we have studied more complex surface charge structures, combining both aspects discussed above: a non-vanishing net charge of the wall and small-scale heterogeneities of the surface charge distribution
	(see Fig. \ref{fig:chargePatterns}). First, we have found a way to fine tune the behavior of the fluid with respect to the transition between the linear and the non-linear response regime by adjusting the area fraction
	of the charged surface and thus effectively tuning the net charge of the wall (see Fig. \ref{fig:allCharge}). Second, we have found a clear dependence of the decay behavior of the electrostatic potential on the lateral
	wavelength of the surface charge structure, where longer wavelengths translate into a longer-ranged decay of the potential away from the wall (see Fig. \ref{fig:allPots}). This effect, as well as all other behaviors of the decay of the potential
	found in the present study can readily be understood on the basis of analytic predictions obtained in the previous study in Ref. \cite{Mussotter2018}, where a connection between the lateral wavelengths and the normal decay behavior has been derived
	(see Eq. \eqref{eq:sinProp}). We note, that within the linear regime the predictions of this previous study provide excellent agreements with the present results, despite its much more simplistic fluid description.
	Finally, we compared the surface charge distributions discussed in the present study in terms of the surface contribution to the grand potential. This confirmed the different influences of the average surface charge as well as of the wavelength of the actual surface charge distribution. Since the strength of the surface contributions is linked directly to the solubility of the corresponding surfaces, our analysis also provides insights into the solubility of large particles carrying a surface charge.

	In conclusion, the present study displays a powerful and very flexible approach to study the effect on the density profiles and the electrostatic potential in contact with surfaces with a broad range of possible surface charge heterogeneities.
	The fully three-dimensional results reveal a strong sensitivity on the overall charge as well as on the detailed shape of the surface pattern.

	Building on previous, more simplistic fluid descriptions \cite{Mussotter2018}, this framework still can be extended in various ways in order to incorporate more realistic and sophisticated models.
	First, much more elaborate density functionals have already been used to account for even more reliable fluid descriptions. These provide a starting point for further extending of the present analysis.
	For example, analyzing equal particle sizes and low ionic strengths heavily narrows the range of occurrence of important effects, where, e.g., Ref. \cite{Podgornik2016} shows possible ways for studies beyond these restrictions.
	Second, in the present study, no wetting or bulk phase transitions have been investigated. In the future, the study of such transitions and their influence on the fluid behavior appears to be promising.
	Finally, the present study has been restricted to periodic surface charge patterns. This is solely done for the sake of simplicity. The investigation of random, disordered surface charge distributions is likely to lead to
	further interesting effects.

	\section*{Acknowledgement}
	We acknowledge a remark by an anonymous referee concerning the solubility of large particles.

\section*{Data availability statement}
	The data that support the findings of this study are available from the corresponding author upon reasonable request.
	\appendix
	\section{Details of the discretization of the system \label{Sec:AppDisc}}
	In order to tackle the situation described in Sec.~\ref{Sec:Formalism}, the system of size $L_x\times L_y\times L_z$ is divided in $N_x$, $N_y$, and $N_z$ cells in the respective directions of space. Each cell is
	of size $|\mathcal{C^*}| = \Delta_x^*\times\Delta_y^*\times\Delta_z^*$ with
	\begin{equation}
		\Delta_i^* = \frac{L_i}{N_i} = \frac{P_i R_1}{N_i},
	\end{equation}
	where $\Delta_i^*$ is the resolution of the numeric calculations with $i\in\{x,y,z\}$, and $P_i$ is the length in units of the particle radius $R_1$. Consequently, $\Delta_i = \frac{P_i}{N_i}$ is the dimensionless resolution and $|\mathcal{C}|=\Delta_x\times\Delta_y\times\Delta_z$ the dimensionless cell size.
	With this division into cells, the position dependence can be captured by indices, which denote the respective cell. E.g., $\rho^{(\alpha)}_{i,j,k}$ describes the density of particle species $\alpha\in\{1,2,3\}$ in the cell located
at the interval $(i\Delta_x,(i+1)\Delta_x]\times(j\Delta_y,(j+1)\Delta_y]\times(k\Delta_z,(k+1)\Delta_z]$. We note that --- in contrast to $q,\vec{\rho}=(\rho_1,\rho_2,\rho_3),\text{ and }\epsilon_r$ --- the electrostatic potential $\Psi$
is defined at the corners of the cells, i.e., $\Psi_{i,j,k}$ is the electrostatic potential at the point $(i\Delta_x,j\Delta_y,k\Delta_z)$.
This leads to the discrete version of the Euler-Lagrange equations (see Eq. \eqref{eq:ELGDENS})
\begin{align}
	\ln\varphi^{(\alpha)}_{i,j,k} = \mu_{\alpha}^*-\sum_{a,b,c,\beta}p_{\beta}\frac{\partial\Phi}{\partial n^{(\beta)}_{a,b,c}}\omega^{(\beta)}_{\alpha}(a-i,b-j,c-k)+\frac{\partial\beta\mathcal{E}}{\partial\varphi^{(\alpha)}_{i,j,k}}.
\end{align}
Here, $\varphi^{(\alpha)}=|\mathcal{C^*}|\rho_{\alpha}$ is the dimensionless density of species $\alpha$, $\mu_{\alpha}^* = \beta\mu_{\alpha} - \ln(\Lambda_{\alpha}^3/|\mathcal{C}^*|)$ is the corresponding dimensionless effective chemical potential, and the prefactor is $p_{\beta}=-1$ for vectorial
weights $\vec{\omega}^{(\beta)}$ and $p_{\beta}=1$ for scalar weights. The three terms result from the ideal gas contribution, the FMT contribution, and the electrostatic interactions, respectively.

\section{Derivation of the expression for the electrostatic field energy \label{Sec:ApUel}}
As illustrated, e.g., in Ref.~\cite{Jackson2014}, one possible way of determining the equilibrium form $\Psi(\vec{r})$ of $\psi(\vec{r})$ in Eq. \eqref{eq:betaU} is a variational approach. Along these
lines we introduce 
\begin{equation}
	\mathcal{E}[\psi,q,\epsilon_r,\sigma] = \Int{\mathcal{V}}{3}{r}\left(\frac{\epsilon_0\epsilon_r(\vec{r})}{2}(\nabla\psi(\vec{r}))^2-q(\vec{r})\psi(\vec{r})\right)-\Int{\partial\mathcal{V}}{2}{s}\sigma(\vec{s})\psi(\vec{s},0),
	\label{eq:AppendixEdef}
\end{equation}
where $q(\vec{r})=e(\rho_2(\vec{r})-\rho_3(\vec{r}))$ is the local charge density, and $\sigma(\vec{s})$ is the surface charge density at the wall $\partial\mathcal{V}=\{\vec{r}\in\mathbb{R}^3|\vec{r} = (\vec{s},z=0)=(x,y,0)\}$.
Furthermore, the equilibrium distribution of the electrical potential $\Psi$ has to fulfill the Poisson equation
\begin{equation}
	\nabla(-\epsilon_0\epsilon_r(\vec{r})\nabla\Psi(\vec{r}))= q(\vec{r})
	\label{eq:Poisson}
\end{equation}
with the boundary condition corresponding to the slope at the wall. This is represented by
\begin{equation}
	\epsilon_0\epsilon_r(\vec{s},0)\vec{n}(\vec{s},0)\cdot\nabla\Psi(\vec{s},0) = \sigma(\vec{s}),
	\label{eq:WallSlope}
\end{equation}
where $\vec{n}(\vec{s},0) =-\vec{e}_z$ is the outer normal vector at $\vec{r}=(\vec{s},0)$. The boundary condition corresponding to a homogeneous bulk system far from the wall is represented by
\begin{equation}
	\Psi(\vec{s},\infty) = 0.
\end{equation}
Provided the correct potential $\Psi$ has been found, $\mathcal{E}$ can be rewritten as
\begin{widetext}
	\begin{align}
		\mathcal{E}[\Psi,q,\epsilon_r,\sigma] =& \Int{\mathcal{V}}{3}{r}\left(\frac{\epsilon_0\epsilon_r(\vec{r})}{2}(\nabla\Psi(\vec{r}))^2-q(\vec{r})\Psi(\vec{r})\right)-\Int{\partial\mathcal{V}}{2}{s}\sigma(\vec{s})\Psi(\vec{s},0)\nonumber\\
		\overset{\eqref{eq:Poisson}}{=}&\Int{\mathcal{V}}{3}{r}\left(\frac{\epsilon_0\epsilon_r(\vec{r})}{2}(\nabla\Psi(\vec{r}))^2-\nabla(-\epsilon_0\epsilon_r(\vec{r})\nabla\Psi(\vec{r}))\Psi(\vec{r})\right)-\Int{\partial\mathcal{V}}{2}{s}\sigma(\vec{s})\Psi(\vec{s},0)\nonumber\\
		\overset{\text{p.i.}}{=}&\Int{\mathcal{V}}{3}{r}\left(-\frac{\epsilon_0\epsilon_r(\vec{r})}{2}(\nabla\Psi(\vec{r}))^2\right)-\Int{\partial\mathcal{V}}{2}{s}\left(\epsilon_0\epsilon_r(\vec{s},0)\vec{n}(\vec{s},0)\cdot\nabla\Psi(\vec{s},0)\right)\Psi(\vec{s},0)\nonumber\\
				 & - \Int{\partial\mathcal{V}}{2}{s}\sigma(\vec{s})\Psi(\vec{s},0)\nonumber\\
		\overset{\eqref{eq:WallSlope}}{=}&\Int{\mathcal{V}}{3}{r}\left(-\frac{\epsilon_0\epsilon_r(\vec{r})}{2}(\nabla\Psi(\vec{r}))^2\right),\label{eq:EUndUelConnection}
	\end{align}
\end{widetext}
leading to 
\begin{equation}
	\beta  U_{\text{el.}}[\vec{\rho}] = -\beta\mathcal{E}[\Psi,q,\epsilon_r,\sigma].
	\label{eq:appendixUE}
\end{equation}
Therefore, given the correct potential $\Psi(\vec{r})$ for the given charge distribution $q(\vec{r})$, the electrostatic contribution to the density functional
can be expressed via $\mathcal{E}[\Psi,q,\epsilon_r,\sigma]$.

\section{Minimization of the auxiliary functional $\mathcal{E}$ \label{Sec:ApE}}
The auxiliary functional $\mathcal{E}$, which is introduced in Eqs. \eqref{eq:Edef} and \eqref{eq:AppendixEdef}, respectively, is constructed in a way, that
its variation with respect to the electrostatic potential $\psi$ is vanishing for the equilibrium potential distribution $\psi=\Psi$ due to the Poisson equation \eqref{eq:Poisson} and its boundary conditions:
\begin{widetext}
	\begin{align}
		\delta\mathcal{E} = & \Int{\mathcal{V}}{3}{r}\left(\epsilon_0\epsilon_r(\vec{r})\nabla\psi(\vec{r})(\nabla\delta\psi) - q(\vec{r})\delta\psi\right)-\Int{\partial\mathcal{V}}{2}{s}\sigma(\vec{s})\delta\psi+\mathcal{O}(\delta q,\delta\epsilon_r)\nonumber\\
		= & \Int{\mathcal{V}}{3}{r}\left(\nabla(\epsilon_0\epsilon_r(\vec{r})\nabla\psi(\vec{r})\delta\psi) - \epsilon_0\nabla\cdot(\epsilon_r(\vec{r})\nabla\psi)\delta\psi - q(\vec{r})\delta\psi\right)-\Int{\partial\mathcal{V}}{2}{s}\sigma(\vec{s})\delta\psi+\dots\nonumber\\
		=& \Int{\mathcal{V}}{3}{r}\left((-\epsilon_0\nabla\cdot(\epsilon_r(\vec{r})\nabla\psi)-q(\vec{r}))\delta\psi\right) + \Int{\partial\mathcal{V}}{2}{s}\left((\epsilon_0\epsilon_r(\vec{s},0)\vec{n}(\vec{s},0)\cdot\nabla\psi - \sigma(\vec{s}))\delta\psi\right)\nonumber\\
		&+\dots\nonumber\\
		\underset{\eqref{eq:WallSlope}}{\overset{\eqref{eq:Poisson}}{=}}& 0\times \delta\psi +\mathcal{O}(\delta q,\delta\epsilon_r).
	\end{align}
\end{widetext}
From this it follows, that for a given and fixed distribution of particles and therefore for a given and fixed charge distribution $q$ and permittivity $\epsilon_r$, the minimum of $\mathcal{E}$ is reached
for $\psi = \Psi$, i.e., the equilibrium potential can be found by a simple minimization of $\mathcal{E}$.
This in turn leads to three types of Euler-Lagrange equations, depending on the distance from the wall, which can be rewritten as
\begin{widetext}
	\begin{align}
		\Psi_{i,j,0} =&\left[2\left(\frac{1}{\Delta_x^2}+\frac{1}{\Delta_y^2}+\frac{1}{\Delta_z^2}\right)\sum_{\alpha,\beta\in\{0,1\}}\epsilon_{r;i-\alpha,j-\beta,0}\right]^{-1}\cdot\left(\sum_{\alpha,\beta\in\{0,1\}}\epsilon_{r;i-\alpha,j-\beta,0}\left[\frac{1}{\Delta_x^2}\left[2\Psi_{i+1-2\alpha,j,0}\right.\right.\right.\nonumber\\
		+&\left.\left.\left. (\Psi_{i+1-2\alpha,j+1-2\beta,0}-\Psi_{i,j+1-2\beta,0}) + (\Psi_{i+1-2\alpha,j,1}-\Psi_{i,j,1}) + \frac{1}{2}(\Psi_{i+1-2\alpha,j+1-2\beta,1}\right.\right.\right.\nonumber\\
		-&\left.\left.\left.\Psi_{i,j+1-2\beta,1})\right]+\frac{1}{\Delta_y^2}\left[2\Psi_{i,j+1-2\beta,0}+(\Psi_{i+1-2\alpha,j+1-2\beta,0}-\Psi_{i+1-2\alpha,j,0})\right.\right.\right.\label{eq:Psi0}\\	
		+&\left.\left.\left. (\Psi_{i,j+1-2\beta,1}-\Psi_{i,j,1})+\frac{1}{2}(\Psi_{i+1-2\alpha,j+1-2\beta,1}-\Psi_{i+1-2\alpha,j,1})\right]+\frac{1}{\Delta_z^2}\left[2\Psi_{i,j,1} + (\Psi_{i+1-2\alpha,j,1}\right.\right.\right.\nonumber\\
		-&\left.\left.\left. \Psi_{i+1-2\alpha,j,0})+(\Psi_{i,j+1-2\beta,1} - \Psi_{i,j+1-2\beta,0}) + \frac{1}{2}(\Psi_{i+1-2\alpha,j+1-2\beta,1}-\Psi_{i+1-2\alpha,j+1-2\beta,0})\right]\right]\right.\nonumber\\
		+&\left.\frac{\chi}{|\mathcal{C}|}\sum_{\alpha,\beta\in\{0,1\}}q_{i-\alpha,j-\beta,0}+2\frac{\chi}{|\mathcal{C}|}\sum_{\alpha,\beta\in\{0,1\}}\sigma_{i-\alpha,j-\beta}\right),\nonumber
	\end{align}
	\begin{align}
		\Psi_{i,j,0<k<N_z} =&\left[2\left(\frac{1}{\Delta_x^2}+\frac{1}{\Delta_y^2}+\frac{1}{\Delta_z^2}\right)\sum_{\alpha,\beta,\gamma\in\{0,1\}}\epsilon_{r;i-\alpha,j-\beta,k-\gamma}\right]^{-1}\cdot\left(\sum_{\alpha,\beta,\gamma\in\{0,1\}}\epsilon_{r;i-\alpha,j-\beta,k-\gamma}\cdot\right.\nonumber\\
				    &\left.\left[\frac{1}{\Delta_x^2}\left[2\Psi_{i+1-2\alpha,j,k}+(\Psi_{i+1-2\alpha,j+1-2\beta,k}-\Psi_{i,j+1-2\beta,k}) + (\Psi_{i+1-2\alpha,j,k+1-2\gamma}\right.\right.\right.\nonumber\\
		-&\left.\left.\left.\Psi_{i,j,k+1-2\gamma})+\frac{1}{2}(\Psi_{i+1-2\alpha,j+1-2\beta,k+1-2\gamma}-\Psi_{i,j+1-2\beta,k+1-2\gamma})\right]+\frac{1}{\Delta_y^2}\left[2\Psi_{i,j+1-2\beta,k}\right.\right.\right.\nonumber\\
		+&\left.\left.\left.(\Psi_{i+1-2\alpha,j+1-2\beta,k}-\Psi_{i+1-2\alpha,j,k}) + (\Psi_{i,j+1-2\beta,k+1-2\gamma}-\Psi_{i,j,k+1-2\gamma})\right.\right.\right.\label{eq:Psik}\\
		+&\left.\left.\left.\frac{1}{2}(\Psi_{i+1-2\alpha,j+1-2\beta,k+1-2\gamma}-\Psi_{i+1-2\alpha,j,k+1-2\gamma})\right]+\frac{1}{\Delta_z^2}\left[2\Psi_{i,j,k+1-2\gamma}\right.\right.\right.\nonumber\\
		+& \left.\left.\left.(\Psi_{i+1-2\alpha,j,k+1-2\gamma} - \Psi_{i+1-2\alpha,j,k})+(\Psi_{i,j+1-2\beta,k+1-2\gamma} - \Psi_{i,j+1-2\beta,k}) \right.\right.\right.\nonumber\\
		+&\left.\left.\left. \frac{1}{2}(\Psi_{i+1-2\alpha,j+1-2\beta,k+1-2\gamma}-\Psi_{i+1-2\alpha,j+1-2\beta,k})\right]\right]+\frac{\chi}{|\mathcal{C}|}\sum_{\alpha,\beta,\gamma\in\{0,1\}}q_{i-\alpha,j-\beta,k-\gamma}\right),\nonumber
	\end{align}
	\begin{align}
		\Psi_{i,j,N_z}=&\left[2\left[\left(\frac{1}{\Delta_x^2}+\frac{1}{\Delta_y^2}+\frac{1}{\Delta_z^2}\right) + \frac{\kappa R_1}{\Delta_z}\right]\sum_{\alpha,\beta\in\{0,1\}}\epsilon_{r;i-\alpha,j-\beta,N_z-1}\right]^{-1}\hspace*{-0.3cm}\cdot\left(\sum_{\alpha,\beta\in\{0,1\}}\epsilon_{r;i-\alpha,j-\beta,N_z-1}\cdot\right.\nonumber\\
			&\left.\left[\frac{1}{\Delta_x^2}\left[2\Psi_{i+1-2\alpha,j,N_z}+(\Psi_{i+1-2\alpha,j+1-2\beta,N_z}-\Psi_{i,j+1-2\beta,N_z}) + (\Psi_{i+1-2\alpha,j,N_z-1}\right.\right.\right.\nonumber\\
		-&\left.\left.\left. \Psi_{i,j,N_z-1})+\frac{1}{2}(\Psi_{i+1-2\alpha,j+1-2\beta,N_z-1}-\Psi_{i,j+1-2\beta,N_z-1})\right]+\frac{1}{\Delta^2_y}\left[2\Psi_{i,j+1-2\beta,N_z}\right.\right.\right.\nonumber\\
		+&\left.\left.\left.(\Psi_{i+1-2\alpha,j+1-2\beta,N_z}-\Psi_{i+1-2\alpha,j,N_z}) + (\Psi_{i,j+1-2\beta,N_z-1}-\Psi_{i,j,N_z-1})\right.\right.\right.\label{eq:PsiN}\\
		+&\left.\left.\left.\frac{1}{2}(\Psi_{i+1-2\alpha,j+1-2\beta,N_z-1}-\Psi_{i+1-2\alpha,j,N_z-1})\right]+\frac{1}{\Delta^2_z}\left[2\Psi_{i,j,N_z-1} + (\Psi_{i+1-2\alpha,j,N_z-1}\right.\right.\right.\nonumber\\
		-&\left.\left.\left. \Psi_{i+1-2\alpha,j,N_z})+(\Psi_{i,j+1-2\beta,N_z-1}-\Psi_{i,j+1-2\beta,N_z})\right.\right.\right.\nonumber\\
		+&\left.\left.\left. \frac{1}{2}(\Psi_{i+1-2\alpha,j+1-2\beta,N_z-1}-\Psi_{i+1-2\alpha,j+1-2\beta,N_z})\right]\right]+\frac{\chi}{|\mathcal{C}|}\sum_{\alpha,\beta\in\{0,1\}}q_{i-\alpha,j-\beta,N_z-1}\right.\nonumber\\
		-& \left.\frac{\kappa R_1}{\Delta_z}\sum_{\alpha,\beta\in\{0,1\}}\epsilon_{r;i-\alpha,j-\beta,N_z-1}(\Psi_{i+1-2\alpha,j,N_z} + \Psi_{i,j+1-2\beta,N_z} + \frac{1}{2}\Psi_{i+1-2\alpha,j+1-2\beta,N_z})\right)\nonumber.
	\end{align}
\end{widetext}
Besides the dimensionless resolutions $\Delta_i$, the dimensionless cell volume $|\mathcal{C}|$, the solvent particle radius $R_1$, and the Debye length $1/\kappa$, the parameter $\chi = \frac{9\pi l_\mathrm{B}}{R_1}$ with the Bjerrum length $l_{\mathrm{B}}$ is introduced here.

\end{document}